\documentclass[psfig]{mn2e}
\def\ltsima{$\; \buildrel < \over \sim \;$}
\def\simlt{\lower.5ex\hbox{\ltsima}}
\def\gtsima{$\; \buildrel > \over \sim \;$}
\def\simgt{\lower.5ex\hbox{\gtsima}}
\input{epsf}

\title{Simultaneous ram pressure and tidal stripping; how dwarf spheroidals
lost their gas}

  \author[Mayer et al.]
{Lucio Mayer $^1$, Chiara Mastropietro$^1$, James Wadsley$^2$,
Joachim Stadel  $^1$ 
\newauthor{\& Ben Moore }$^1$
\\$^1$Institute of Theoretical Physics,
University of Z\"urich, Winterthurerstrasse 190, 8057 Zurich, Switzerland
\\$^2$Department of Physics \& Astronomy, McMaster University, 1280 Main St.
West, Hamilton ON L8S 4M1 Canada}
\begin{document}

%\pagerange{\pageref{firstpage}--\pageref{lastpage}} \pubyear{00}

\maketitle

\label{firstpage}

\begin{abstract}

We perform high-resolution N-Body+SPH simulations of gas-rich 
dwarf galaxy satellites orbiting 
within a Milky Way-sized halo and study for the first time the combined effects
of tides and ram pressure. The structure of the galaxy models and the orbital
configurations are chosen in accordance to those expected in a $\Lambda$CDM 
Universe.
%Simpler numerical experiments which include only ram
%pressure or tidal stripping are also performed.
While tidal stirring of disky dwarfs produces objects whose stellar
structure and kinematics resembles that of dwarf spheroidals after a few
orbits, ram pressure 
stripping is needed to entirely remove their gas component.
Gravitational tides can aid ram pressure stripping by 
diminishing the overall potential of the dwarf, but tides also
induce bar formation which funnels gas inwards making subsequent
stripping more difficult.
This inflow is particularly 
effective when the gas can cool radiatively. 
Assuming a low density of the hot Galactic corona consistent with
observational constraints, dwarfs with  $V_{peak} < 30$ km/s can be 
completely stripped of their gas content on orbits with pericenters of
50 kpc or less. Instead, dwarfs with more massive dark haloes and 
$V_{peak} > 30$ km/s lose most
or all of their gas content only if a heating source keeps the gas
extended, partially counteracting the bar-driven inflow. We show that 
the ionizing radiation from the cosmic UV background at $z > 2$ can
provide the required heating. 
In these objects most of the gas is removed or becomes ionized at the
first pericenter passage, explaining the early truncation of the star
formation observed in Draco and Ursa Minor.
Galaxies on orbits with larger pericenters and/or falling into the Milky
Way halo at lower redshift can retain significant amounts of the 
centrally concentrated gas. These dwarfs would continue to form stars
over a longer period of time, especially close to pericenter passages,
as observed in Fornax and other dSphs of the Local Group.
The stripped gas breaks up into individual clouds pressure confined by the 
outer gaseous medium that have masses, sizes and densities comparable to 
the HI clouds recently discovered around M31.

\end{abstract}


\section{Introduction}

The origin of dwarf spheroidal galaxies (dSphs) is a long standing subject
of debate. These galaxies are gas poor or completely devoid of gas and are
the faintest galaxies known. Objects matching the definition of dSphs
are found both in galaxy clusters and in groups, but it is certainly the 
Local Group the place where dSphs have been better studied 
thanks to their proximity. 

\begin{figure*}
\hskip 1truecm
\epsfxsize=14truecm 
\epsfbox{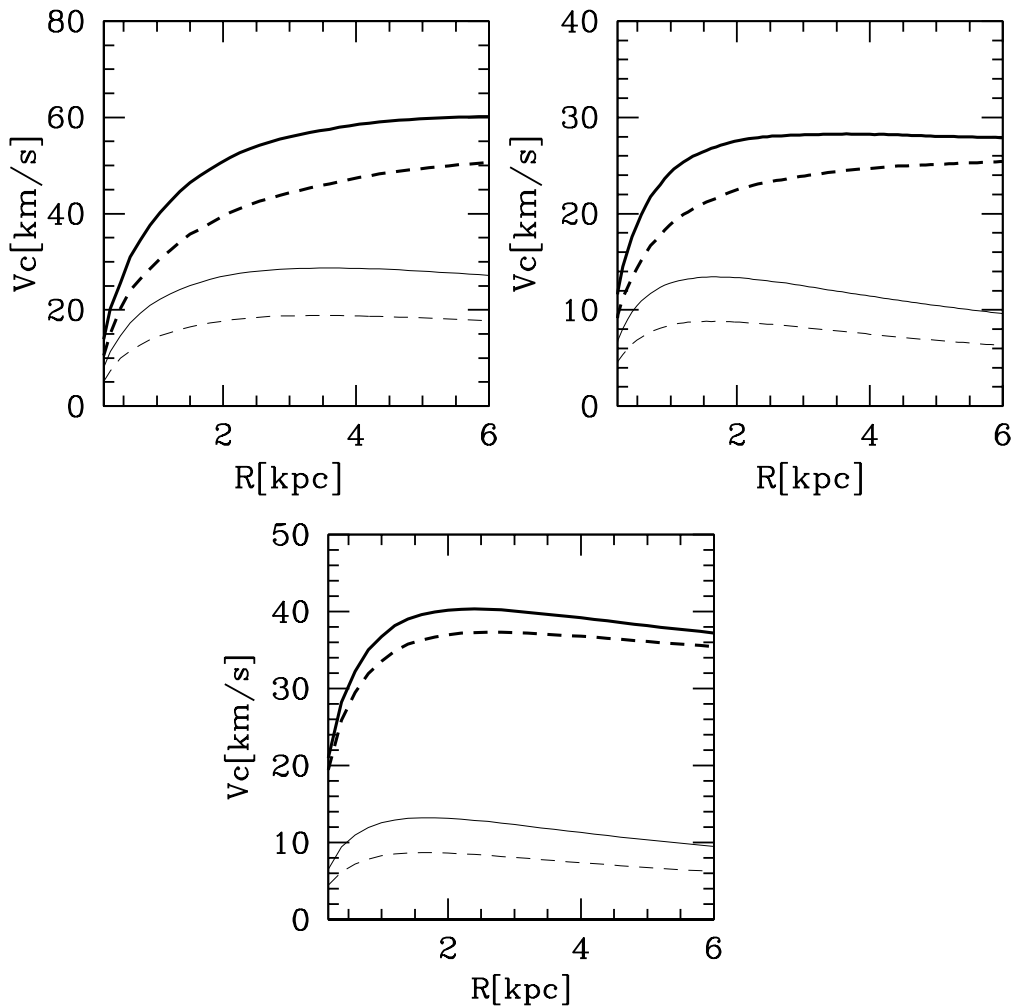}
%\epsfbox{rcurve1c4.ps}
%\epsfxsize=7truecm 
%\epsfbox{rcurve2c4.ps}
%\epsfxsize=7truecm
%\epsfbox{rcurvec20.ps}
\caption{Initial rotation curves of the dwarf galaxy models.The lines show,
respectively, the total rotation curve (thick solid), the contribution
of dark matter (thick dashed), that of stars (thin solid) and that of gas
(thin dashed). From top left to bottom, models V60c4, V30c4 and V40c20 
are shown.}
\end{figure*}

Two main formation paths have been proposed
explain their origin, nature of nurture. In one scenario these faint
galaxies could be the result of cosmic reionization and stellar 
feedback quenching gas accretion and star formation in low mass haloes
(Dekel \& Silk 1986; Thoul \& Weinberg 1996;
Bullock, Kravtsov \& Weinberg 2000; Ferrara \& Tolstoy 2000
; Benson et al. 2002a, 2002b; Somerville 2002; Susa \& Umemura 2004). 
However, there is no obvious signature of reionization in the star formation
history of these galaxies (Grebel \& Gallagher 2004), which in many cases
continued to form stars over more than 10 Gyr, and for some dSphs the
most recent inferred halo masses are likely too large ($10^8-10^9 M_{\odot}$, 
see Kleyna et al. 2002; Wilkinson et al. 2004;Kazantzidis et al. 2004)
for supernovae winds to
remove most of the baryons (Mori, Ferrara \& Madau 2002; Mayer \& Moore 
2004; Read \& Gilmore 2004).

In addition, these models
do not naturally explain why dSphs share some structural similarities
with equally faint but gas-rich, much younger disk-like galaxies, dwarf 
irregulars (dIrrs), notably their exponential light distribution.
Furthermore they do not give rise to the morphology-density relation observed
in groups and clusters, namely the fact that dSphs tend to be concentrated
towards the central galaxies while dIrrs are found at much larger distances,
nor they explain the dichotomy between dominance of rotation for the stars of
most dIrrs and dominance of pressure support for those of dSphs.
Various environmental
effects have been scrutinized to explain similarities and
differences between the two types of galaxies postulating that dSphs
are somehow transformed dIrrs. Ram pressure
stripping of gas from dwarf irregular-like progenitors caused by
a hot diffuse gaseous corona surrounding the Milky Way
has been proposed as a way to 
explain the low gas contents of dSphs together with the similar stellar light 
distributions of the two types of dwarfs (Einasto et al. 1974; 
Faber \& Lin 1983; Grebel, Gallagher \& Harbeck 2003), and has also
been considered as the origin of HI clouds possibly associated 
with some dSphs and transitional dIrrs/dSphs like Phoenix 
(Blitz \&  Robishaw 2000;
Gallart et al. 2001). Van der Bergh (1996) 
has argued that the
correlation between galactocentric distance and gas content in dSphs
is a signature of ram pressure stripping. Tidal stripping of dwarf irregulars
has  also been considered as an alternative way of removing gas from
the dwarfs (e.g. Ferguson \& Binggeli 1994).

Recently, high
resolution N-Body/SPH simulations were used to show that tides do not
only remove gas and stars 
but can also reshape the stellar components of a dwarf
disky galaxy resembling a dwarf irregular (Mayer et al. 2001a,b). This
``tidal stirring'' can turn a rotationally supported disk into
a pressure supported spheroidal stellar system by means of bar formation
and a subsequent buckling instability. 
The mechanism is particularly effective for satellites of massive galaxies
that are strongly tidally shocked owing to the plunging orbits expected 
in CDM models and works for 
systems having a variety of dark halo profiles, including cuspy profiles
(Mayer et al. 2002).  The transformation is not necessarily
associated with intense tidal mass loss; indeed, objects with very
high-mass-to light ratios, thus embedded in dense, massive dark halos, 
can still be reshaped from the instabilities induced by tidal shocks 
albeit losing minimal mass (Mayer et al. 2001b).

A problem of this model, however, is that 
tides cannot remove completely the gas from the disky progenitors of dSphs; 
the low gas fractions of dSphs can be explained by only invoking rapid 
consumption from star formation of the
remaining gas in a series of tidally triggered bursts associated
with bar-driven gas inflows (Mayer et al. 2001b).
However, although
intermittent star formation is observed in some dSphs (Gallart
et al. 1999; Hernandez, Gilmore \& Valls-Gabaud
2000), the star formation rates needed to completely
consume the gas are well in excess
of those implied by reconstructions of the star formation
histories of dSphs using detailed color-magnitude diagrams 
(Mayer et al. 2001b). 
%This problem is particularly severe 
%for satellites with dense and massive halos like Draco (Kleyna
%et al. 2002; Wilkinson et al. 2004) 
%that would undergo minimal stripping.
Moreover, Draco and a few other dSphs do not show extended star 
formation histories, instead seem to have ceased star formation about 
10 Gyr ago (Grebel et al. 2004).

\begin{figure*}
\hskip 1.3truecm
\epsfxsize=14truecm
\epsfbox{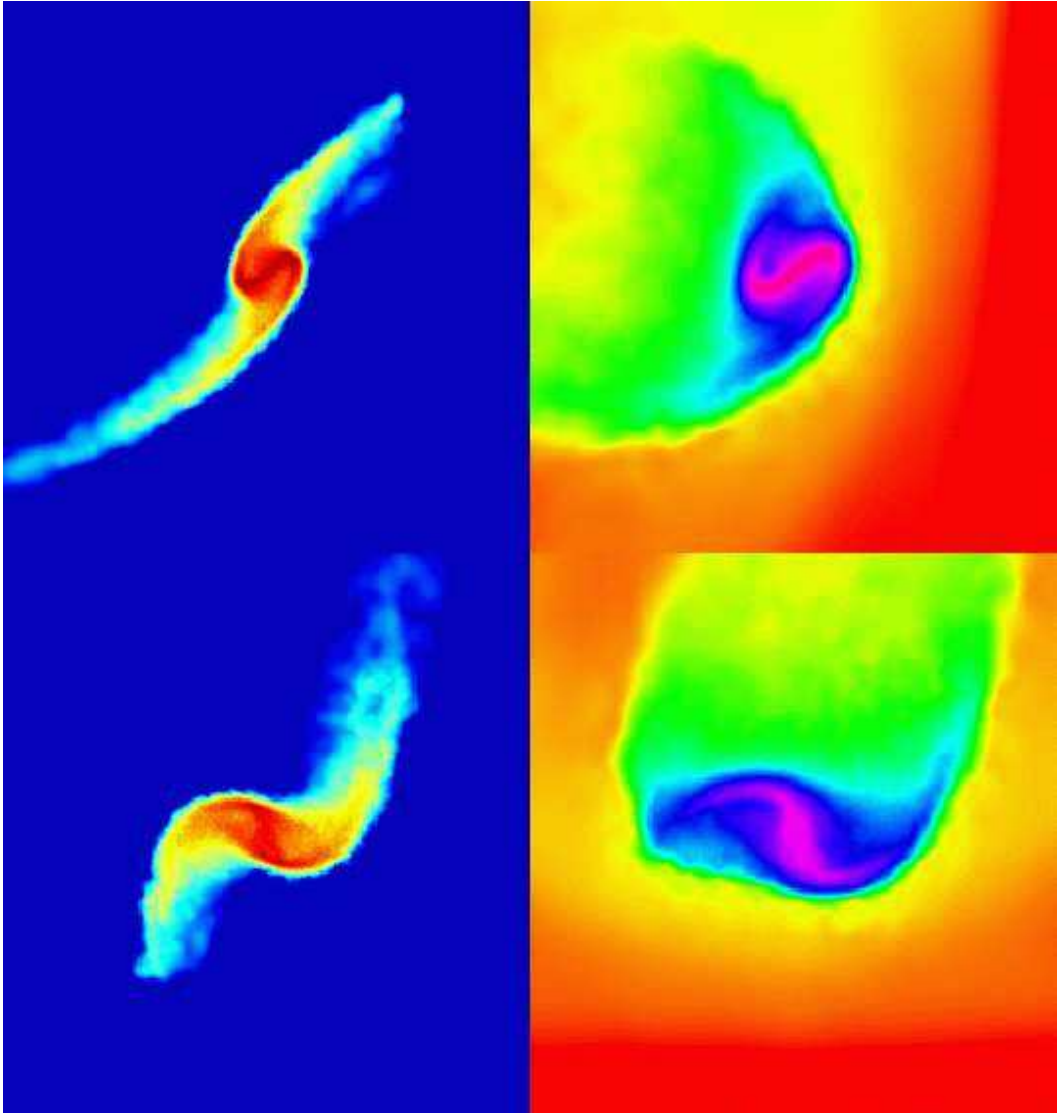}
\caption{Color coded logarithmic density maps for run Rc4aADHR. 
Projections
  along (top) and perpendicular to the orbital plane (bottom) are shown for
  the stars (left) and for the gas component (right). Snapshots are taken
at $T=1.7$ Gyr, while crossing pericenter for the first time. Boxes are
100 kpc on a side for the stars and 30 kpc on a side for the gas (a smaller
box is used for the gas to show its structure in greater detail). The two 
color maps are such that colors go, respectively, from blue to red through 
yellow (stars) and from red to magenta through yellow and green (gas) 
going from lower to higher densities.}
\end{figure*}

It is likely that both gravitational tides and hydrodynamical processes
play a role (Gavazzi et al. 2001).
For example, ram pressure could
act simultaneously with tides
and increase the efficiency of gas removal (Mayer \& Wadsley 2003).
Recent 2D and 3D grid-based hydrodynamical simulations of ram pressure 
stripping of disk-like dwarfs in
poor groups show that ram pressure alone can remove most of the gas in dwarf
galaxies with halo circular velocities lower than $30$ km/s (Marcolini, 
Brighenti \& D'Ercole 2003). 
However, these and other simulations (Mori \& Burkert 2000)
do not use models of dwarf galaxies
directly based on the $\Lambda$
CDM model and do not include the effect of tides.
%induced instabilities and tidal stripping. More in general, the models adopted
%in this and other works (e.g. Mori \& Burkert 2000 on dwarf galaxies
%moving through the intracluster medium) do not model the self-gravity
%of all galaxy components and therefore cannot account for the full
%hydrodynamical and gravitational response of the system.
%As we shall see later in this paper, the impact that tides have on the 
%effectiveness of ram pressure is non trivial especially because of the 
%way a fully self-gravitating disk responds to the tidal disturbance 
%reorganizing its structure. 
The effectiveness of both ram pressure and tidal stirring will also 
depend on the orbital
history of the dwarfs, and this is where the cosmological framework comes 
into play. 
For example dwarfs that fell into
the primary halo at very high redshift likely had orbits with smaller 
pericenters relative to those falling later on (Mayer et al. 2001b), thus
suffering stronger tidal shocks, and possibly also stronger
ram pressure in a denser gaseous corona, while being 
exposed to the effect of the cosmic UV background (Bullock,
Kravtsov \& Weinberg 2000; Somerville 2002; Barkana \& Loeb 1999;
 Shaviv \& Dekel 2003).
Indeed cosmological simulations of galaxy formation incorporate all such
mechanisms but their resolution is still too coarse to allow a study of 
their effects at the scale of the tiny dwarf spheroidals 
(Governato et al. 2004). 

In this paper we will study for the first time the
combined effects of ram pressure and tidal stirring 
using high resolution three dimensional N-Body+SPH simulations
of dwarf galaxies orbiting within the Milky Way dark halo and gaseous
corona. We will use galaxy models and orbits consistent 
with the predictions of $\Lambda$CDM simulations. The simulations 
include also radiative cooling as well as heating and ionization from the
cosmic UV background radiation.
The main goal of the present work will be to establish whether these two 
mechanisms can explain the present-day low gas content of dSphs if these
galaxies are the descendants of gas-rich dwarfs with structure similar
to today's dIrrs.

\begin{table*}
\centering
\caption{Parameters of the full interaction (FI) simulations.
Column 1: Name of model
Column 2: Peak circular velocity of the satellite (km/s)
Column 3: Satellite halo concentration
Column 4: Disk scale length  (kpc)
Column 5: Apocenter distance (kpc)
Column 6: Pericenter distance (kpc)
Column 7: Thermal physics (AD=adiabatic, RC=radiative
cooling, RCUV=radiative cooling + cosmic UV background). All runs include
tides and ram pressure except those referred to as TD which do not include
ram pressure.
Column 8: name of run (HR stands for high resolution, see text).
Column 9: initial gas mass (in units of $10^6 M_{\odot}$)
Column 10: final gas mass (in units of $10^6 M_{\odot}$)
}
\begin{tabular}{lcccccccccc}
Model & $V_{peak}$ & $c$ & $R_{h}$ & $r_{apo}$ & $r_{peri}$ & physics &
Run & $M_{gas_i}$ & $M_{gas_f}$\\
\\
V60c4 & 62  & 4 & 2.2 & 250 &  50 & AD & Rc4aAD & 450 & 55 &  \\
V60c4 & 62   & 4 & 2.2 & 250 &  50 & AD & Rc4aADHR & 450 & 60 & \\
V60c4 & 62   & 4 & 2.2 & 250 &  50 & RC & Rc4aRCHR & 450 &  380 & \\
V60c4 & 62   & 4 & 2.2 & 250 &  50 & RCUV & Rc4aRCUV & 450 & 290 & \\
V28c4 & 25  & 4 & 1.1 & 250 &  50 & TD(AD) & Rc4bTD  & 45 & 20 & \\
V28c4 & 25   & 4 & 1.1 & 250 &  50 & AD & Rc4bAD & 45 & 0 & \\
V28c4 & 25   & 4 & 1.1 & 250 &  50 & RC & Rc4bRC & 45 & 9.5 & \\
V28c4 & 25   & 4 & 1.1 & 150 &  30 & RC & Rc4cRC & 45 & 0 & \\
V40c20  & 42   & 20 & 0.5 & 150 &  30 & TD(AD) & Rc20TD & 46.5 & 22.2 & \\
V40c20 & 42 &  20 & 0.5 & 150 &  30 & AD & Rc20AD & 46.5 & 0 &  \\
V40c20   & 42 & 20 & 0.5 & 150 &  30 & RC & Rc20RC & 46.5 & 3.55 \\
V40c20  & 42 & 20 & 0.5 & 150 &  30 & RCUV & Rc20RCUV & 46.5 & 0.65 \\
V40c20  & 42 & 20 & 0.5 & 250 &  50 & AD & Rc20bAD & 46.5 & 0 & \\
V40c20  & 42 & 20 & 0.5 & 250 &  50 & RC & Rc20bRC & 46.5 & 10 & \\

%\hline

\end{tabular}
%\label{t:simul}
\end{table*}

\section {Initial Conditions}

\subsection{Galaxy models}

Models comprise an exponential disk of gas and stars embedded in a live NFW 
halo for the dwarf, and a live NFW halo plus hot gas distribution with the 
same density profile for the primary galaxy. 
Halo virial parameters (mass, radius, 
circular velocity) are consistent with the LCDM model. Although we believe 
that the Milky
Way halo has a virial velocity $V_{vir} = 140-160$ km/s 
(Klypin, Zhao \& Somerville 2002) we use a 
slightly larger value, $V_{vir}=180$ km/s, in order to have 
$V_{peak} = 230$ km/s at about 10 kpc from the center without
including a disk component. We choose a concentration  
$c=10$ for the MW halo, this being typical for LCDM halos at a scale of 
$\sim 10^{12} M_{\odot}$ (Bullock et al. 2001). With our choice
of parameters the shape of the
rotation curve is quite flat out to $100$ kpc and the resulting halo mass 
within $100$ kpc is 
$\sim 8 \times 10^{11} M_{\odot}$, consistent with 
the orbital dynamics of the Magellanic Clouds and distant dSphs 
(Lin, Jones \& Klemola 1995; 
Dehnen \& Binney 1998). Overall our halo model resembles model A3
in Klypin, Zhao \& Somerville (2002), which matches simultaneously 
most of the known observational constraints within the framework of the 
LCDM model.
The hot gas is in hydrostatic
equilibrium within the halo potential and its temperature is $\sim 10^6$ K
at 50 kpc, consistent with OVI, OVII and X-ray absorption measurements in 
the Galactic halo (Sembach et al.2003). 
 Assuming an isotropic model, the halo temperature
at a given radius $r$ is determined by the cumulative mass distribution $M(r)$ of
the dark, stellar and gaseous components of the MW beyond $r$ and by the density
profile $\rho_h(r)$ of the hot gas: 
\begin{equation}\label{gashalo} 
T(r) = \frac{m_p}{k_B} \frac{1}{\rho_h(r)} \int_{r}^{\infty} \rho_h(r)\frac{GM(r)}{r^2} \, dr \, , 
\end{equation} where $m_p$ is the proton mass, $G$
and $k_B$ are the gravitational and Boltzmann constants.
Its density at 50 kpc from the center is $8 \times
10^{-5}$ atoms/cm$^3$ (see also Mastropietro et al. 2004).
At a distance of $\sim 100$ kpc or more, where the dwarf galaxies that
we have simulated spend most of the time while orbiting
within the primary halo (see below), the hot halo density is 
$< 10^{-5}$ atoms/cm$^3$.
Such a hot diffuse halo is also a natural 
expectation of the current LCDM paradigm of structure formation (White \&
Frenk 1991; Governato et al. 2004; Sommer-Larsen, Portinari \& Goetz 2003; 
Maller \& Bullock 2004).
%More details about how the model was constructed are given in Mastropietro et
%al. 2004. 

The multi-component dwarf galaxy models are built using the technique developed
by Hernquist (1993) and later refined by Springel \& White (1999) (for further
details on the modeling see also Mayer et al. 2002).
They comprise an exponential disk of gas and stars embedded in an NFW halo. 
We used three dwarf models with $V_{peak}$ $=28$, $42$ or $62$ km/s
km/s and concentration $c=4$ (the first two models) and $c=20$ (the third
model). 
%Note that the halos in these models have virial velocities $V_{vir} 
%=20$ (the first two models) and $50$ km/s (the third model);
%$V_{peak}$ can be significantly higher than $V_{vir}$ because of a 
%combination of halo concentration, addition of baryonic disk mass and 
%adiabatic contraction of the halo in response to such addition, all included in%our models.  Here the main reason is
%halo concentration since, as it will be explained in a moment, the dwarfs have fairly low disk mass fractions. 
The choice of halo concentration can be
quite flexible since cosmological simulations show a large scatter of
such property at small mass scales, mostly due to a large spread in halo 
formation times (Bullock et al. 2001). 
The rotation curves of the models are reported in Figure 1; while
the rotation curves of models V60c4 and V30c4
are slowly rising in agreement with those of  todays' dwarf irregulars, 
that of model
V40c20 rises quite steeply owing to the much higher halo concentration.
A steep rotation curve is a more appropriate choice for the progenitors of 
Draco or Ursa Minor that have very high central dark matter
densities (see Mayer et al. 2001b; Hogan \& Dalcanton 2000).

Lokas (2002) and Kazantzidis et al. (2004)
 have shown that dSphs 
kinematics suggests they are embedded in halos with $V_{peak} \sim 20-35$
km/s.  We used even larger values of $V_{peak}$ in our initial conditions
since these can later decrease as a result of tidal mass loss, by roughly 
a factor of 2 according to the recent cosmological 
simulations by Kravtsov, Gnedin \& Klypin (2004). 
%which corresponds to a reduction of total mass 
%within $r_{v_{peak}$} of about a factor of 10.

\begin{figure*}
\hskip 1.3truecm
\epsfxsize=14truecm
\epsfbox{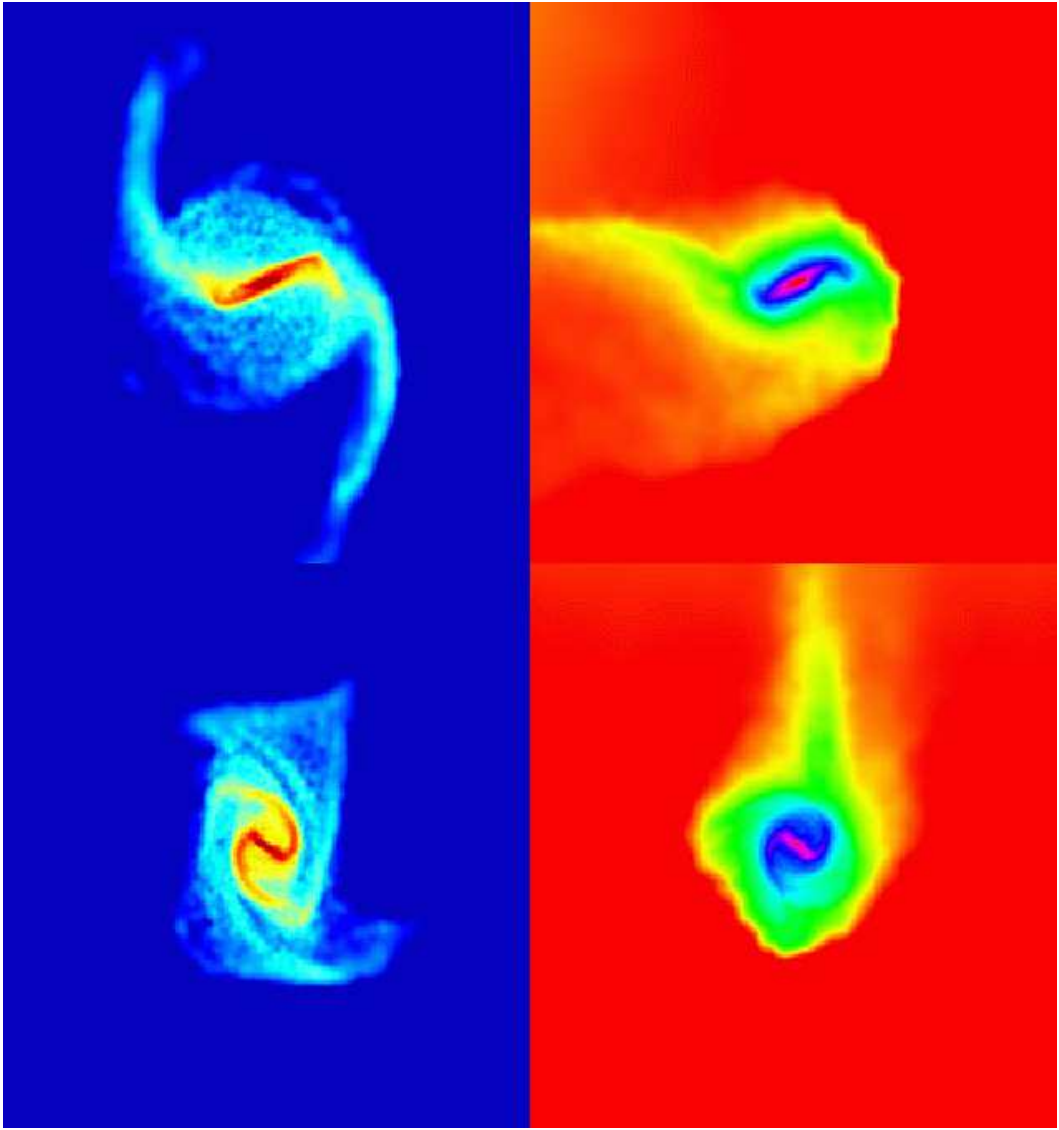}
\caption{Color coded logarithmic density maps for run Rc4aADHR. 
Projections along (top)
and perpendicular to the orbital plane (bottom) are shown for the stars
(left) and for the gas component (right). Snapshots are taken
at $T=3$ Gyr, after first pericenter passage. Boxes are 50 kpc on a side.}
\end{figure*}

The galaxy models have disks of mass $\approx 4\%$ of the total mass, 
lower than the
universal baryon fraction but quite typical for present-day LSB or dIrr galaxies
(Jimenez et al. 2003). Following, Mo, Mao \& White (1998), the 
exponential scale length $R_h$ of the disk is determined by the spin parameter and
concentration $c$ of the halo; we use $\lambda=0.043$ for models Rc4a and Rc4b, 
and $\lambda=0.08$ for model Rc20a,b (the mean spin of cosmological halos is 
$\lambda \sim 0.05$, see e.g. Gardner 2001, and 20\% of halos have spins as
high as $0.1$ due to the log-normal shape of the probability distribution of halo spins).
The disk sizes in models  V28c4 and V40c20 are
the same (the higher spin parameter compensates for the higher halo
concentration) and, since also the disk mass is the same by choice,
we can isolate the role of halo concentration (or, equivalently, $V_{peak}$)
by comparing the two models. Note that all the models have a central 
surface brightness of $23$ mag arcsec
$^{-2}$ in the B band for a stellar-mass-to light ratio of 2, typical of
dIrrs and dwarf spirals (see de Blok \& McGaugh 1997).
Galaxies with these low disk mass fractions are stable to bar formation
in isolation irrespective on halo concentration (Mayer \& Wadsley 2004).
However they can become bar unstable when tidally
perturbed by a much more massive halo. 

We use $3 \times 10^5$  particles for the dark halos
of the dwarfs and $3 \times 10^4$ particles for both the stellar 
and the gaseous disk of the dwarf.
The Milky Way halo is modeled with $10^6$ 
dark matter particles, while $5 \times 10^5$ particles are used for the 
superimposed gaseous halo (the mass of hot halo particles is
$8 \times 10^4 M_{\odot}$). Each simulation has a total of 
$\sim 1.8 \times 10^6$ particles, which makes them quite computationally
demanding. Two runs employing model V60c4
were repeated with increased resolution in the disk of the dwarf,
($2 \times 10^5$ particles for both its stellar and 
its gaseous component, see Table 1). 
The choice of putting a large fraction of the total number of particles in the 
primary halo is motivated by the desire of minimizing spurious two-body heating between
halo particles and those of the disks of the dwarfs (Moore, Katz \& Lake
1996; Mayer 2004).
%Also, the high number of gas particles in the primary halo is required to 
%avoid artificial stripping of the dwarf due to two-body scattering between
%gas particles or widely different masses; with our choice the mass of
%gas particles in the primary and in the dwarf is indeed the same, which
%was shown to be the ideal choice in Mastropietro et al. (2004).
We use a fairly large gravitational softening for the gas and
dark matter particles in the primary system, $\varepsilon= 2$ kpc to minimize
discreteness noise in the potential.
 For the dwarf models instead,
the softening is  $\varepsilon= 0.3 R_h$ for halo particles 
and  $\varepsilon= 0.1 R_h$ for star and gas particles, where $R_h$ is the
disk scale length (see Table 1). 
% We recall that
%by scaling the softening with the peak velocity we essentially scale it with
%the disk scale length ($V_{max}$ is reached at about $2.2$ scale-lengths in
%an exponential disk), hence with the typical scale length of the baryons
%in the dwarf.

\subsection{Simulations}

The dwarf galaxy satellites are placed on bound eccentric orbits in the
primary halo. In Mayer et al. (2001b) we explored a wide range of orbital
eccentricities and initial disk inclinations. Tidally induced 
transformation was found to be essentially independent on disk inclination
(with only purely retrograde encounters versus purely prograde encounters
showing  significant
differences). We also found that the orbital time, and not the orbital
eccentricity, is the most sensitive
orbital parameter for causing the transformation since it  
determines both the strength and number of tidal shocks. Here we choose orbits
with an apocenter to pericenter ratio of 5, corresponding to the mean value 
for satellites in cosmological simulations (Ghigna et al. 1998, Gill et
al. 2004), and we consider apocenters of 250 or 150 kpc 
(the pericenter being 50 or 30 kpc) corresponding to 
orbital times of, respectively, 3.2 and 1.3 Gyr. The dwarf galaxies start
at apocenter and are evolved for 3 to 5 orbits with their disks initially
inclined 60 degrees with respect to their orbital plane. 
Different mean orbital
distances radii could be associated to satellites infalling into 
the Milky Way halo at 
different cosmic epochs (Mayer et al. 2001b; Kravtsov et al. 2004; Gill
et al. 2004), say at $z > 1$ or $z < 1$, for, respectively, the smaller 
and larger orbital time.

The simulations were performed with the parallel Tree+SPH code GASOLINE  
(Wadsley, Stadel \&  Quinn 2004).
The internal
energy of the gas is integrated using the asymmetric formulation, that gives
results very close to the entropy conserving formulation (Springel
\& Hernquist 2003) but conserves
energy better. In simulations without
radiative cooling the gas can heat or cool only by adiabatic compression
or expansion. Dissipation in shocks is modeled using the quadratic term
of the standard Monaghan artificial viscosity. The Balsara correction term
is used to reduce unwanted shear viscosity (Balsara 1995).
A detailed description of the SPH code is given in Wadsley et al. 2004. 

\begin{table*}
\centering
\caption{Parameters of the wind tunnel (WT) simulations.
Column 1: Model Galaxy (see Table 1).
Column 2: Gas physics (AD=adiabatic, RC=radiative cooling).
Column 3: Angle between z-axis of the disk and direction of the flow.
Column 4: Mass remaining in the disk at after $5 \times 10^8$ years (in units
of $10^7 M_{\odot}$).
Column 5: Stripping radius in kpc.
Column 6: Stripping radius (in kpc) predicted using analytical prediction (see text).
Column 7: name of run (LR stands for ``low resolution'', see text).
}
\begin{tabular}{lcccccccccc}
Model  &  gas physics  & $i$ & $M_{fin}$ & $R_{strip_1}$ & $R_{strip_2}$ & 
RUN \\
\\ 
V28c4  & AD &  0 & 0 &  0 & 0 & T1c4a &\\
%V28c4   & AD &  0 & 50 &  0 & 0 & T1c4aLR &\\
V28c4   & AD &  90  &  0  &  0   & 0   &    T2c4a    & \\
V28c4   & RC & 0 & 0 &  0 &  0 &  T3c4a &\\
V28c4   & RC & 45 & 0 & 0 & 0 & T4c4a & \\
V28c4   & RC & 90 & 6 &  0 & 0 & T5c4a & \\
V60c4   & AD & 0 & 57 &  4.25 & 5.8 &  T1c4b & \\
V60c4   & AD & 30 & 66 &  4.25 & 5.8 & T2c4b & \\
V60c4   & AD & 60 & 64 &  4.25 & 5.8 & T3c4b & \\
V40c20   & AD & 0 & 4.38 &  1.58 & 1.6 & T1c20 & \\
V40c20   & AD & 0 & 3.6 &  1.5 & 1.6 & T1c20LR & \\
V40c20   & RC & 0 & 6.2 &  1.75 & 1.6 & T2c20 & \\
V40c20   & RC & 90 & 7.85  & 2 & 1.6 & T3c20 & \\

%\hline

\end{tabular}
%\label{t:simul}
\end{table*}

We perform both adiabatic runs and runs with radiative cooling. 
We use a standard cooling function for a 
primordial mixture of 
hydrogen and helium (the metallicity in dSphs is indeed much lower than
solar, with $-1 <$ <Fe/H> $ <
-2$, see Gallagher, Grebel \& Harbeck 2003).  
%By construction
Cooling shuts off below $10^4$ K. The initial temperature
of the gaseous disks is $8000$ K for the most massive model ($V_{max}=62$ km/s)
and slightly lower for the other lighter models. 
%We tested by evolving
%the models in isolation that with such choice of the temperature the gas disk
%is roughly in hydrostatic equilibrium with the gravitational potential
%out to a few disk scale lengths (since we use
%a single temperature disk a flaring of the disk at the
%outermost radii is unavoidable).  
%In adiabatic runs the gas can heat
%by compression and shocks but can cool only by expanding.
The purpose of running adiabatic simulations is two-fold; first, it
helps in understanding the interplay between thermodynamical and
mechanical effects of ram pressure as we shall see in the following
section,
and second heating by external radiation fields or stellar feedback, especially in low mass objects as those considered in this paper, might overtake 
radiative cooling and create a hot phase that would essentially behave as
an adiabatic gas (Thacker \& Couchman 2000; Marcolini et al. 2003;
Springel \& Hernquist 2003).  Therefore radiative
cooling and adiabatic runs can be regarded as two extreme cases
crudely bracketing the conditions of the real interstellar medium in which a
hot and a cold phase will coexist at all times.
Finally, in a few runs with cooling we also include a uniform external 
radiation field which models the cosmic UV background arising at high redshift.
This is implemented using the Haardt \& Madau (1996) model which
includes photoionizing and photoheating rates produced by QSOs and galaxies
starting at $z=7$. We implicitly assume that the dwarf enters 
the Milky Way halo at $z=7$. The ionizing flux remains quite high until $z=2$, 
therefore the dwarf is exposed to the radiation for a duration of
$\sim 5$ Gyr.
%This is a conservative choice since newer models
%consistent with the optical depth to polarization of CMB radiation detected by
%WMAP as well as other observational constraints predict UV fluxes that are
%much higher than those used here already at $z=15$ owing to the effect
%of Population III stars (Choudhuri \& Ferrara 2004). 

\begin{figure*}
\hskip 1.3truecm
\epsfxsize=14truecm
\epsfbox{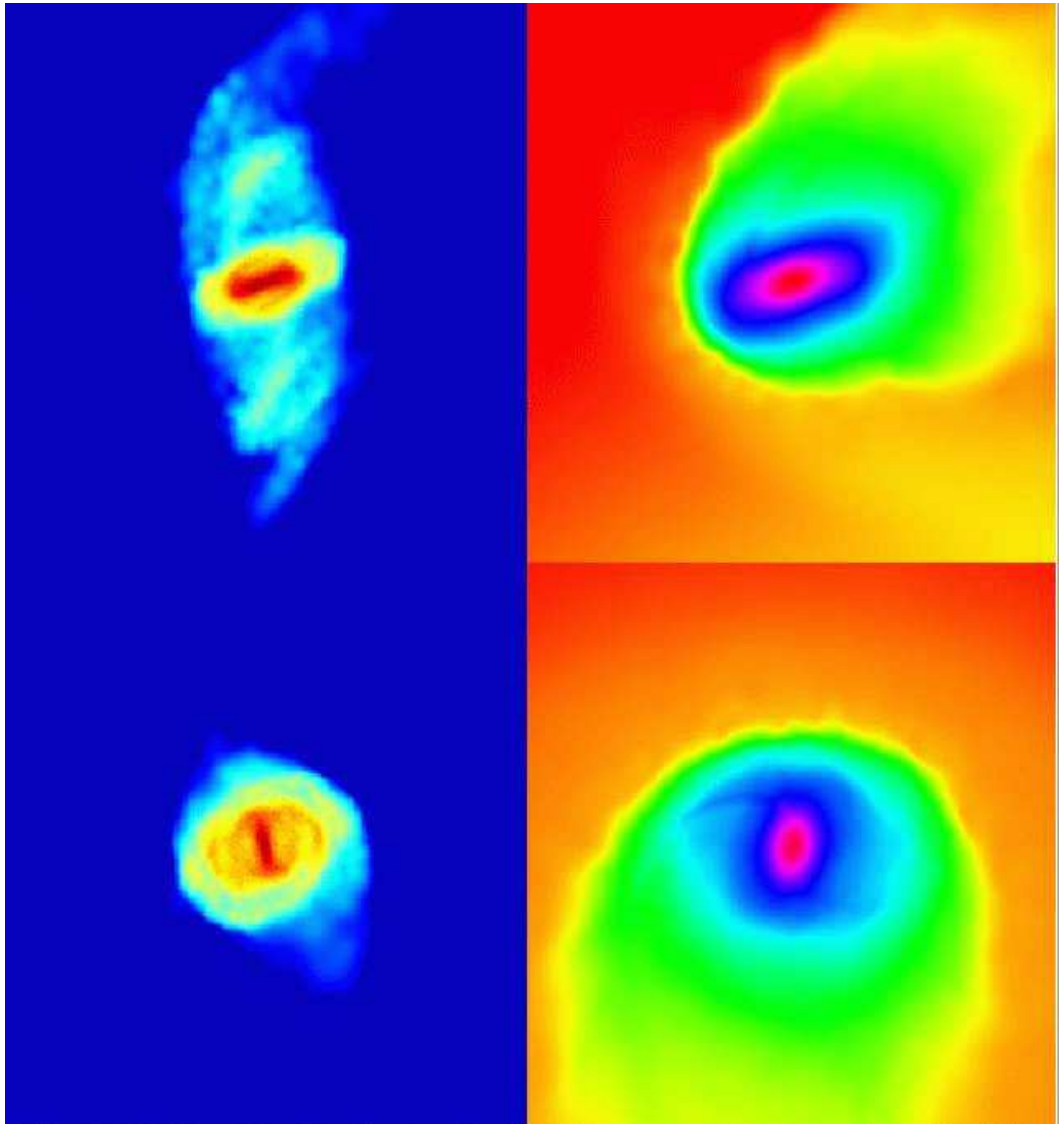}
\caption{Color coded logarithmic density maps for run Rc4aADHR. 
Projections along (top)
and perpendicular to the orbital plane (bottom) are shown for the stars
(left) and for the gas component (right). Snapshots are taken
at $T=6$ Gyr, after the second pericenter passage. Boxes are 50 kpc on a side
for the stars and 25 kpc on a side for the gas.}
\end{figure*}

%Ram pressure stripping is believed
%to be quite sensitive to the relative inclination between the disk and its
%direction of motion through the ambient medium; such angle changes
%with time in the simulations since the orbits are not straight line paths. 
We complement our ``full interaction'' runs (hereafter FI 
runs), so called because they include both tides and ram pressure,
with wind tunnel simulations aimed at studying the effect of ram 
pressure alone.
%In these simulations the dwarf models feel
%the ram pressure from a costant wind.  
The volume
of the simulations is a tube with rectangular section, having a base equal 
to the diameter of the dwarf model (out to its virial radius) and
height $h=vt$, where $v$ is the typical velocity of the satellite at 
pericenter in the FI runs
($290 \,\textrm{km}\,\textrm{s}^{-1}$) and $t$ is set $\sim 0.5$ Gyrs ($\simgt
 t_{enc}=R_{peri}/V_{peri}$, where $t_{enc}$ is the time the dwarf spends near
pericenter of the orbits in the FI runs).
We represent the hot gas as a flux of
particles moving with a velocity $v$ along the longest axis of the tube, this
being parallel or at an angle with the angular momentum of the disk depending
on the run. The hot particles have an initial random distribution, a
temperature $T=10^6 \textrm{K}$ and a number density $n_h=8\times 10^{-5}
\textrm{cm}^{-3}$. We use periodic boundary conditions in order to restore
the flow of hot gas that leaves the tube.
We will indicate these as WT runs (where WT stands for
wind tunnel).
These experiments allow to resolve the background gas flow with very 
high resolution, thus limiting numerical artifacts
like the spurious stripping induced by transfer of momentum
in two-body collisions between the hot halo
particles and the gas particles in the disk of the dwarf (Abadi, Moore
\& Bower 1999).
%The resolution tests in tube flows conducted by Mastropietro et al. (2004)
%have shown that convergence in the amount of mass stripped is approached when 
%the mass of the particles in the flow is equal to that of the gas particles 
%in the galaxy model. Therefore the setup of the WT runs is such that
%the ratio between the mass of the background gas particles and dwarf 
%gas particles is equal to one. 
The mass of the gas particles is equal to that
used for the dwarfs in the FI runs, i.e. $8 \times 10^4 M_{\odot}$, for the
runs employing our largest dwarf model, V60c4, while it is 8 times
lower for the runs employing the other two models, V28c4 and V40c20, since
these have correspondingly lighter disks. The (adiabatic) runs with models 
V28c4 and V40c20 where repeated with a lower resolution of the background flow,
equal to that of the FI runs.
The total number of particles used in the hi-res WT runs thus varies
from $10^6$ in runs with models  V28c4 and V40c20 to $2 \times 10^5$ in runs 
with model V60c4.
%In these simulations we set the velocity of the galaxy equal to the typical 
%velocity at the pericenter of the orbits in the FI simulations 
%($V_{peri} \sim 290$ km/s on the chosen eccentric orbits). 

\begin{figure*}
\hskip 1.3truecm
\epsfxsize=14truecm
\epsfbox{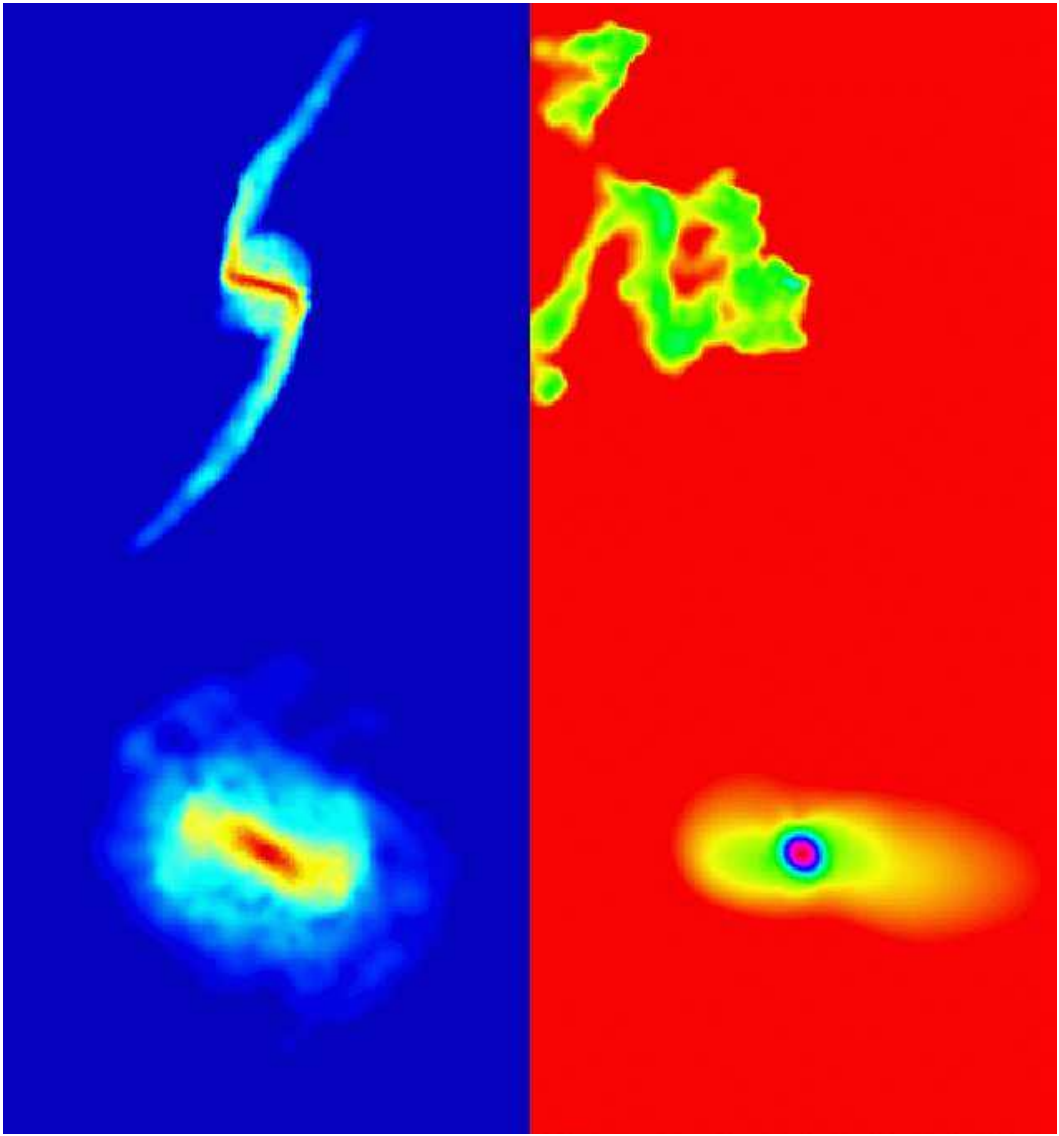}
\caption{Color coded gas density (right) and stellar logarithmic density maps  (left)
for run Rc4bRC. Projections along the orbital plane 
are shown in a box of 30 kpc on a side 
at $T=1.7$ Gyr (top) and in a box of 10 kpc on a side at $t=6$ Gyr (bottom)
after the second pericenter passage. All boxes are centered on the center
of mass of the bound stars. Re-accretion of gas between the two
different times has clearly occurred (the gas distribution is completely
displaced from the stars at  $T=1.7$ Gyr).}
\end{figure*}

%\begin{figure}
%\epsfxsize=6truecm
%\epsfbox{v30c4RCstarden.ps}
%\caption{Color coded stellar density map for run Rc4bRC, three random
%  projections after $T=1.7$ Gyr. Boxes are 30 kpc on a side.}
%\end{figure}

\section{Results}

Table 1 and 2 report the results of the FI and WT runs.
Dwarf galaxies undergo severe stripping owing to the combined action
of tides and ram pressure. The overall mass loss varies
depending on the dwarf galaxy model, in particular the
depth of its potential well, on its orbit and on the gas physics
(the balance between cooling and heating). Dwarfs can retain
$20-50\%$ of their initial gas content or be completely stripped.
Ram pressure stripping occurs very rapidly
as the galaxy approaches pericenter for the first time. A bow shock is 
visible (Figure 2) and reflects the fact that the dwarfs
near pericenter are moving at mildly supersonic velocities in the hot halo
owing to their eccentric orbits (the Mach number is $\sim 1.5$).

The tidal shock at first pericenter passage
deforms the disk to an extent that depends
on how deep is the potential well of the galaxy. 
%within a few kiloparsecs from the center%
Symmetric stellar tidal tails develop but 
but the gas generally trails the galaxy since ram
pressure sweeping dominates (compare gas and stars in Figure 3). 
The dark halo loses most of its mass in all cases and the overall depth of 
the potential well is reduced which increases the efficiency of gas removal.
The stars develop a bar instability due
to the strong tidal perturbation (Figure 3) which drives a radial gas 
inflow (Mayer et al. 2001b). This inflow counteracts
the reduction of the mean restoring force by moving some gas 
deeper towards the center of the dwarf where it is harder to strip
(note the central density concentration in Figure 4-6).
As a result on subsequent orbits ram pressure stripping continues but 
with much lower intensity. 
%$Removal of such highly concentrated gas will then 
%depend on its pressure support (and therefore on the assumed thermodynamics)
%which determines the degree of concentration of its
%radial distribution; the more diffuse the distribution, the easier
%for tides and ram pressure to remove it. 

%Table 1 describes the  FI runs, 
%while Table 2 described the WT runs.
In what follows we will describe the results quantitatively by 
analyzing the gas mass loss in FI and WT runs as well as
the fate of the stripped gas.

\subsection{Gas mass loss}

\subsubsection{Wind tunnel (WT) runs}

%These runs allow to isolate the role of ram pressure 
%since tidal forces are not included. 
Figure 7-10 show the evolution of the
gas component of the dwarfs.
%The galaxies are evolved in the
%tube flows for $5 \times 10^8$ years, a timescale comparable to the time
%galaxies spend in the densest region of the halo in the FI runs (the typical 
%timescale of pericenter crossing is $R_{peri}/V_{peri}=2 \times 10^8$ years 
%in the FI runs). 
In the absence of an external perturbation the stellar
disks are stable to bar formation.
%, a major difference
%compared to the FI runs. 
Removal of gas begins immediately; a bow shock develops
and some gas in the disk rises above the disk plane  
slightly more than $10^7$ years. This is evident in the 
early truncation of the gas surface density profiles (see Figure 12).
The stripped gas
is diffuse in adiabatic runs whereas it breaks into clumps pressure 
confined by the external medium in the cooling runs (see Figure 8).
The evolution of the temperature of the gas is shown in Figure 11.
As the hot halo pushes the gas
out of the disk it also rises its temperature to $\sim 10^5$ K as a result
of compression. If radiative cooling is included the
gas quickly cools down to $10^4$ K; cooling occurs on a timescale shorter
than $10^7$ years since the gas is heated to temperatures at which the cooling
rate is maximum. 
Occasionally gas particles that did not gain enough 
momentum to leave the disk fall back towards the midplane as they cool down 
(compare for example  Figures 7 and 8 for run T2c20 - the disk appears more 
intact at a later time). The distribution
of the gas becomes slightly more concentrated towards
the midplane of the disk (see the evolution of the surface density profiles
in Figure 12), and subsequent stripping is harder. In adiabatic runs, instead, 
the heated gas remains hot and expands; the gas disk thickens and it 
is stripped more easily than in the cooling runs (see Figure 8)
as its own pressure support makes it less bound to 
the disk potential. No stripped gas falls back to the disk.

\begin{figure*}
\hskip 1.3truecm
\epsfxsize=12truecm
\epsfbox{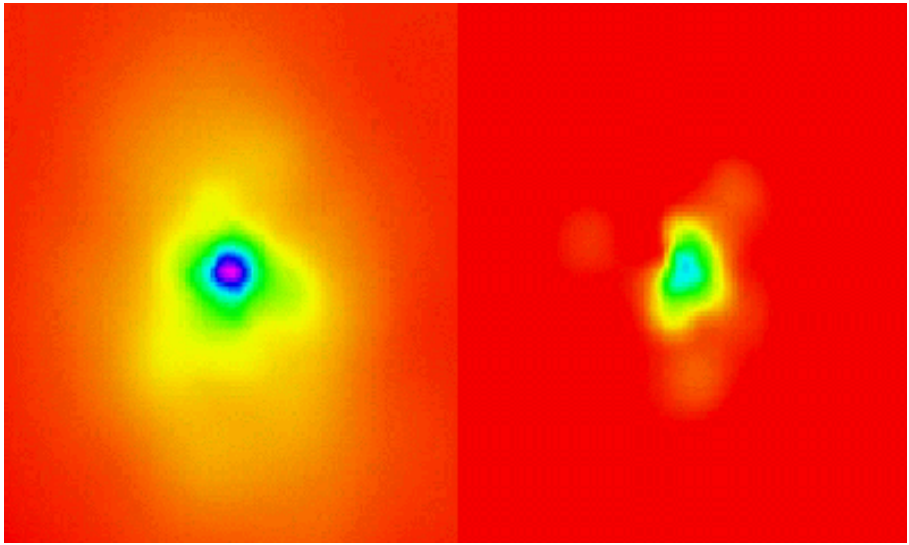}
\caption{Color coded logarithmic gas density maps for run Rc20RC (left) and for run Rc20RCUV (right) at $t=5$ Gyr, after three pericenter passages.
The projection along the orbital plane is shown. The boxes
are 10 kpc on a side and are centered on the center of mass of the bound 
stars. Clearly the gas remaining in the dwarf is much less and much more
diffuse when UV heating is added. }
\end{figure*}

We consider as ``stripped'' all the gas that is removed from the
stellar disk. 
%In principle such gas can still be 
%bound to the halo; we note that it typically leaves the disk with 
%escape velocities of order 
%$100$ km/s and, since this velocity is lower than the escape velocity
%from most of our models, will eventually leave the halo as well (gas that 
%is only a few hundred parsecs above the disk can have moderate velocities 
%however and sometimes is reaccreted in cooling runs). 
Table 2 also includes the stripping radii found in the simulations, 
defined as the minimum distance from the center of the dwarf at which gas
particles are stripped.
%(for edge-on runs stripping tends to be
%quite asymmetric with respect to the disk symmetry axis and we take 
%an average between the two sides of the disk). 
Some of the WT runs were performed twice, using a resolution of the background
flow equal or higher than that of the FI runs employing the same model (see
section 2.2); we found differences within 20\% for the gas mass loss and
even less for the stripping radii.
%A steepening of the gas density profile towards
%the center, however, is
%present in all cases as a result of compression from the external medium
%(Figure 8).
Table 2 reports the gas mass in the disk measured after 
$0.2$ Gyr. The difference in the amount of stripped mass between runs with
and without cooling  is visibly higher than that between runs adopting 
different disk inclinations; indeed inclination effects are appreciable only 
when extreme cases are compared, namely an edge-on versus a face-on run (where
the disk inclination is intended relative to the direction of motion of
the background gas flow).  We note that for the most massive model, V60c4,
we have performed only adiabatic runs since our main goal here 
is to estimate the maximum amount of mass that can be stripped
by ram pressure alone.
The difference between 
runs with and without radiative cooling is smaller in terms of stripping
radii than it could be guessed 
from the difference in the amount of stripped mass (Table 2).
The reason is that the 
distribution of gas in the cooling runs becomes more concentrated (thus a 
given radius contain more mass) as we noted before.

%\begin{figure}
%\epsfxsize=6truecm
%\epsfbox{v30c4RCgasdent130.ps}
%\caption{Color coded gas density map for run Rc4bRC, three random projections
%after $T=6$ Gyr. Boxes are 30 kpc on a side.}
%\end{figure}

%\begin{figure}
%\epsfxsize=6truecm
%\epsfbox{v30c4RCstardent130.ps}
%\caption{Color coded stellar density map for run Rc4bRC, three random
%  projections after $T=6$ Gyr. Boxes are 30 kpc on a side.}
%\end{figure}

\begin{figure}
\hskip 1truecm
\epsfxsize=5.5truecm
\epsfbox{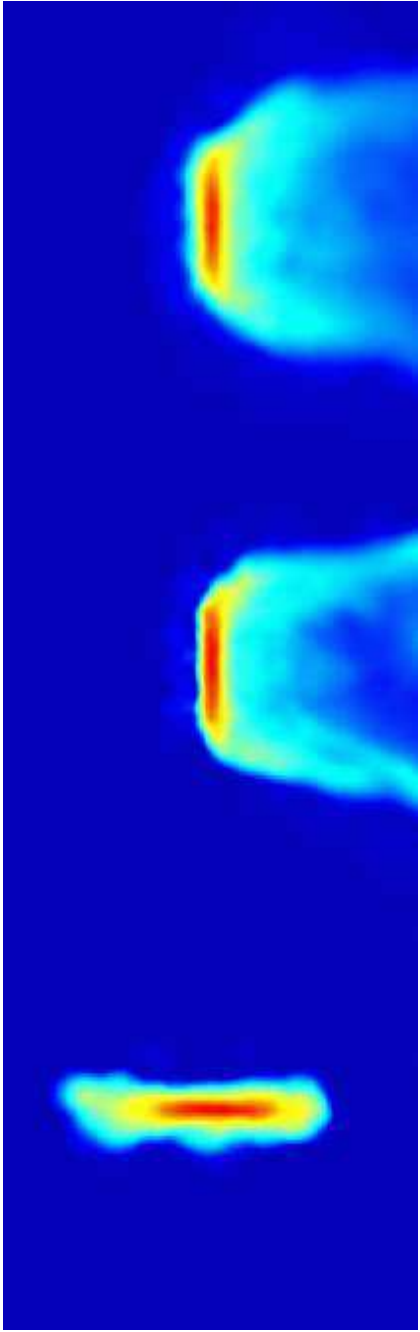}
%\epsfxsize=10truecm
%\epsfbox{tube300c20.ps}
%\epsfbox{tube300.xz.ps}
%\epsfxsize=6truecm
%\epsfbox{tube300c.xz.large.ps}
%\epsfxsize=6truecm
%\epsfbox{tube300cp.large.xy.ps}
\caption{Color coded logarithmic gas density map for three wind tunnel (WT)
runs. From top to
bottom, an edge-on view after $t =0.05$ Gyr of the gas disk in runs T1c20,
T2c20, T3c20. The background gas flows along the horizontal axis from left
to right (from right to left in the bottom panel).
Boxes are 10 kpc on a side.}
\end{figure}

%Note that we have not run all combinations of inclinations and gas
%thermodynamics (adiabatic or cooling) for all the different
%models; each simulation is computationally expensive at the resolution we
%use and therefore we mainly aimed at obtaining a measure of the
%maximum amount of gas that can be stripped for each of the models, which
%is why adiabatic runs and face-on runs were performed for each model
% and 
%(2) extract the general trends as a function of gas physics and 
%disk inclination. 

Our results show that 
model V28c4 is completely stripped in all cases (e.g. Figure 9)
except when cooling and 
an inclination of 90 degrees with respect to the flow are used at the
same time (both choices go in the direction of minimizing stripping).
Instead the 
models V60c4 and V40c20 always retain a significant amount of gas. This 
indicates that if the progenitors of dSphs once had massive halos 
corresponding to  $V_{peak} \sim 50$ km/s, as suggested by 
recent cosmological simulations  (Kravtsov, Gnedin \& Klypin 2004),
ram pressure alone was not able to remove their gas early on unless the
halo density was once significantly higher than $10^{-4}$ atoms cm$^{-3}$. 
On the other end, these results also suggest that efficient
gas removal from ram pressure could have happened at some later stage
if tides were able to weaken the potential well of the dwarf bringing
$V_{peak}$ down to $\sim 30$ km/s. Interestingly, an NFW halo with 
$V_{peak} \sim 30$ km/s or lower does fit the present-day velocity
dispersion of dSphs (Lokas 2002; Kazantzidis et al. 2004). We will come
back to this in the following sections

\begin{figure}
\hskip 1truecm
\epsfxsize=5.5truecm
\epsfbox{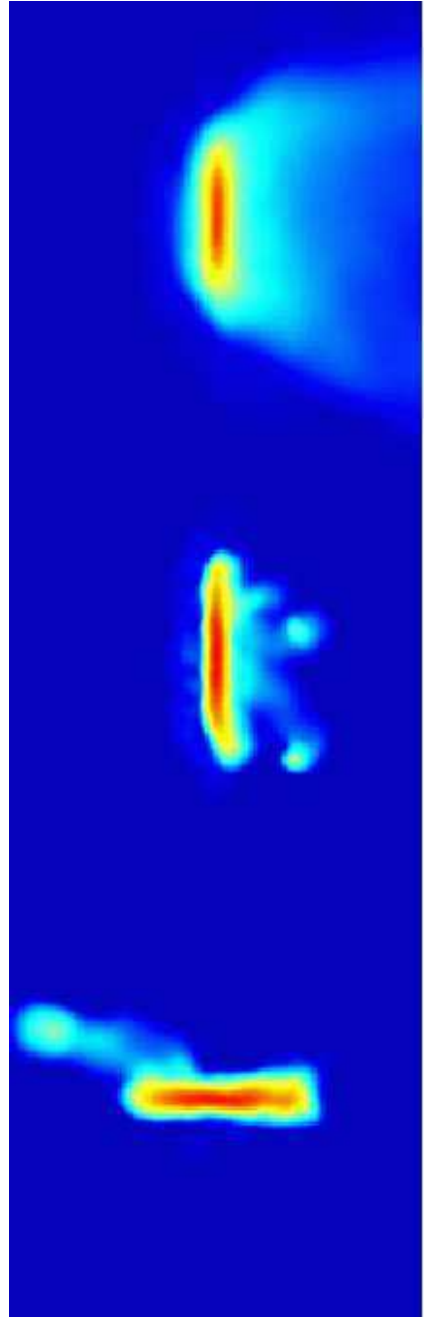}
%\epsfxsize=10truecm
%\epsfbox{tube300c20.ps}
%\epsfbox{tube300.xz.ps}
%\epsfxsize=6truecm
%\epsfbox{tube300c.xz.large.ps}
%\epsfxsize=6truecm
%\epsfbox{tube300cp.large.xy.ps}
\caption{Color coded logarithmic gas density map for three wind tunnel (WT) runs. From top to
bottom, an edge-on view after $t =0.3$ Gyr of the gas disk in runs T1c20,
T2c20, T3c20. The background gas flows along the horizontal axis from left
to right (from right to left in the bottom panel). Boxes are 10 kpc on a side.}
\end{figure}

We tried to compare our numerical results with simple analytical estimates
of the effectiveness of ram pressure stripping.
Gas is removed from the disk of the dwarfs if the ram pressure force
is greater than the restoring force per unit area 
provided by the disk and halo of the dwarf. The condition for ram pressure
stripping for an axisymmetric disk system is expressed by (Gunn \& Gott 1972) 
\begin{equation} 
\rho_h v^2 > 2 \pi G \Sigma(R) \Sigma_g(R), 
\end{equation} where $v$ is the velocity of the galaxy
 with respect to the surrounding medium, $\rho_h$ is the density of the 
hot halo of the MW and $\Sigma_g(r)$ is the cold gas surface density 
at the radius
$R$. $\Sigma$ represents the surface mass density of the
disk. The minimum radius given by Equation (2) is the final stripping radius
$R_{str}$ beyond which the ISM can be removed from the galaxy.
Equation (2), however, neglects the contribution from dark matter to the
restoring force (see Gavazzi et al. 2001). In all of our models the dark 
matter content is high even near the disk; we will simply 
account for that by 
replacing $\Sigma(R)$ with $\Sigma_{dm} + \Sigma_b(R)$, where $\Sigma_{dm}$ is 
the surface mass density of the dark matter in a slab with height 
comparable to the disk thickness and $\Sigma_b(R)$ is the surface mass
density of the baryons within the same slab (in our models the gas component
always gives a non negligible contribution to the disk mass). 
%We note that
%at $z=0$ the contribution of the dark halo to the restoring force
%vanishes since the halo is spherically symmetric. However, most of the
%gas particles will feel the restoring force due to the halo since the are
%located at $z > 0$, and therefore we
%do include   $\Sigma_{dm}$ in our estimate (in principle this choice
%will lead to an overestimate of the restoring force for gas particles
%located very close to the plane $z=0$).  
This formula describes the instantaneous stripping of gas.
We found it difficult to use this prescription as to reflect
both the complex interplay between mechanical and thermodynamical
effects seen in the simulations and the fact that the stripping is
not exactly instantaneous but continues for some time (typically it saturates
after $\sim 0.2$ Gyr).
In particular, $\Sigma_g(r)$ is strongly
time dependent due to the action of heating and cooling as we noted above.
We find that a reasonable prediction can be obtained if we choose not the
initial  $\Sigma_g(r)$ but that at the time the gas begins to heat up
(as little as $10^6$ years after the beginning of the simulations).
The predicted stripping radius (see Table 2) can be
slightly larger than that in the simulations (even for runs with cooling).
However, the prediction is always close 
to the stripping radius measured in the 
simulations at earlier times, i.e. when the approximation of a single, 
instantaneous 
stripping event is better satisfied.

\begin{figure}
\hskip 0.7truecm
\epsfxsize=7truecm
%\epsfbox{tubec4v30CL.ps}
\epsfbox{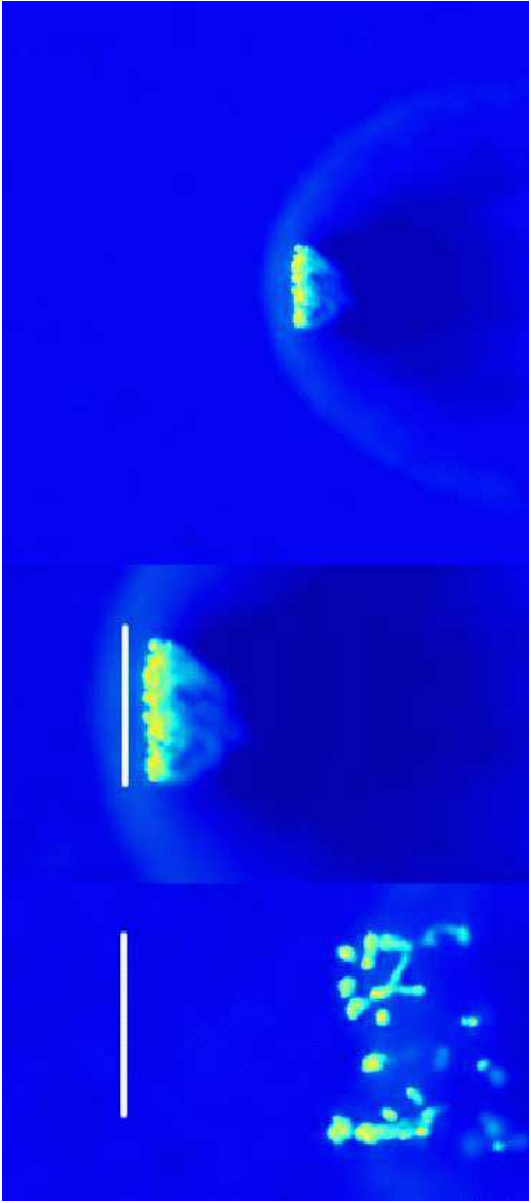}
\caption{Color coded logarithmic gas density map for the gas of the dwarf 
in run
T2c4b. On top the output after $t =0.05$ Gyr is shown in a box of
30 kpc on a side. The bow shock is clearly visible. In the middle
panel we show a zoom-in of the same snapshot (the box is 10 kpc along 
the longest axis). The bottom panel shows a zoom-in of a snapshot taken 
after t=$0.3$ Gyr (the box is also 10 kpc along the longest axis). The white
vertical bar shows the location of the stellar disk. Complete gas stripping 
occurs in this case.
The background gas flows along the horizontal axis from left to right.}
\end{figure}

%Strictly speaking (2) should only be applied
%for face-on stripping whereas only the projection of the ram pressure force
%$\rho_h v^2$ should be used in all cases where the disk symmetry axis does
%not coincide with the direction of the flow. However, several authors have
%shown the differences in mass stripping when the disk is not moving face-on 
%with the flow cannot be simply explained by the projection cosine law (Schulz
%\& Struck 2001;Vollmer et al. 2001). Since a detailed investigation of this
%aspect is beyond the scope of this paper and since we just want to obtain
%a rough prediction for the stripping radius to be compared with the numerical
%results, we use equation (2) without modifications in runs with varying disk 
%inclination.

\subsubsection{Full Interaction (FI) runs}

In these runs the dwarf galaxy evolves under the combined action of tides and 
ram pressure.  The mass loss as a function of time for a number of runs is 
shown in Figure 13.
%As shown in Figure 2 and 4, in addition to an obvious head tail 
%produced by ram pressure, a smaller leading tail is produced by tides.  
At any given time we identify as bound to the dwarf the gas with 
temperature $T < T_{vir}$ (where $T_{vir}$ is the initial virial temperature
of the halo of the dwarf, in the range $\sim 3 \times 10^4 - 10^5 K$ depending
on the model) and which is located within the bound stellar 
component. We checked that by using this definition one essentially
selects the gas that is gravitationally bound to the dwarf.
%comparable to those obtained by tracking the bound mass using the definition
%based on energy balance. With the caveat that one can miss some gas at larger
%radius that is still marginally bound to the system as a whole, this 
%definition has the advantage of being quite close to how gas fractions are
%usually measured in dSphs where one typically searches for neutral or ionized
%gas within the tidal radius of the stellar 
%component (the only difference is thus that we employ the actual 
%bound radius of the dwarf instead of the apparent tidal radius deduced from
%a fit with a King profile). The mass loss as a function of time for a
%number of runs is shown in Figure 16.

%We note that in the FI runs
The angle between the direction of the flow in the hot halo and the plane of 
the disk is continously changing because of the orbit of the galaxy
and since the disk inclination relative
to the orbit changes as the dwarf is tidally torqued by the primary.
We compare the gas mass losses observed in FI runs with those
occurring in the face-on WT runs which yield
the highest possible gas mass loss due to ram pressure only. 
In general galaxies lose more gas compared to
WT runs (compare Table 1 and Table 2 for both adiabatic and cooling runs).
However, as we found previously, only in some of the runs (mostly those
employing model V28c4)
is the dwarf completely stripped of its gas content (see Table 1).
Furthermore, while with tides added a dramatic increase in the
stripping is always observed in adiabatic runs, stripping can be unexpectedly 
less effective in cooling runs (compare run Rc4bRC in Table 1 with all the 
``c4a'' runs in Table 2). 
%, and indeed
%there are cases in which the same dwarf model that was being completely
%stripped in WT runs retains some small amount of gas in FI runs when
%cooling is included (compare run Rc4bRC in Table 1 with all the ``c4a'' runs
%in Table 2). 
This happens because the tidally induced bars
funnel $20-40$\% of the initial gas content of the disk within a radius
smaller than the initial disk scale length of the dwarf. 
%The gas flows in because it is
%tidally torqued by the stellar bar, which robs its angular momentum.
The gas shocks through
the bar as it loses angular momentum and heats up. 
In adiabatic runs this heat
cannot be dissipated quickly and the gas ends up more extended.
%ralenting 
%his fall towards the center by increasing its pressure support.
In cooling runs the gas loses rapidly its thermal energy and there is
not enough pressure to stop its inflow.
As a result gas profiles become much more concentrated in cooling runs
and a larger amount of gas ends up deeper in the potential well where it 
is harder to strip (Figures 14-15). In runs which include the heating from
the UV background the gas remains more extended than with just radiative
cooling (Figure 6) and stripping is much more efficient, with gas mass 
losses closer to those of adiabatic runs.
%This effect is additional to that already seen in WT runs
%(see previous section and Figure 15), namely that the gas that is not removed
%readily cools rapidly after being heated by the external medium and ends
%up more concentrated that at the beginning of the simulation. Hence the
%difference between cooling and adiabatic regimes is amplified in FI runs
%(again compare final masses in Table 1 and Table 2).

\begin{figure}
\hskip 0.7truecm
\epsfxsize=7truecm
\epsfbox{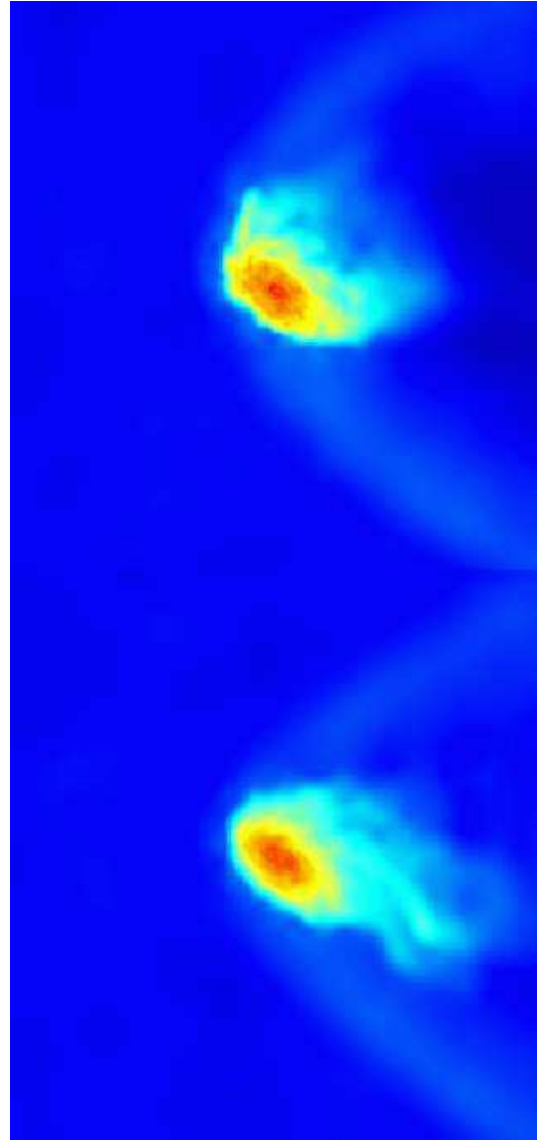}
\caption{Color coded logarithmic gas density map for the gas of the dwarf in 
run T2c4b. From top to bottom, the outputs after $t =0.05$ Gyr and 
$T=0.3$ Gyr are shown. The boxes are 30 kpc on a side. The background flow is
along the horizontal axis from left to right.}
\end{figure}

Despite the complex interplay between bar instability and thermal physics,
it turns out that the fractional amount of stripped mass correlates
well with the initial $V_{peak}$ of the satellite on a given orbit, both 
with and without
cooling (for example compare run Rc4cRC and run Rc20RC in Table 1, or
run Rc4aRCUV and run Rc4bRC).  For the same initial $V_{peak}$, orbits
with smaller pericenter lead to a larger reduction in the $V_{peak}$
due to more numerous and stronger tidal shocks (Figure  16). This
favours ram pressure stripping as it lowers the gravitational restoring
force. Furthermore, on orbits with a pericenter of 30 kpc the maximum halo
gas density is about a factor of 3 higher relative to the case with
a pericenter of 50 kpc, and this also favours ram pressure stripping.
As a consequence gas mass loss is significantly larger and faster in
runs using a more plunging orbit (Table 1).
%The complex interplay between bar instability and cooling is such that
%there is a weak dependence of mass stripping on $V_{peak}$ in FI runs
%with cooling. Runs involving
%$V40c20$ and $c75c4$ are stripped of roughly $80 \%$ of their gas with
%cooling despite the fact that the first model has a lower
%$V_{peak}$ and is evolved on an orbit with
%shorter orbital time, thereby undergoes more pericenter passages. This
%is easy to understand because (1) the bar drives a significant fraction
%of the gas well within the tidal radius of both models for both pericenter
%distances and (2) ram pressure cannot overcome the restoring force
%of gravity once the gas sits so deep in the potential well of the system. 

Estimating the stripping radius from formula (2) is somewhat meaningless in 
these
runs because the depth and shape of the potential of the dwarf 
is continously changing due to the action of tides and the distribution
of the bound gas also changes with time.
Nonetheless, it is noteworthy that the stripping radius estimated using
formula (2) with the parameters of the initial models is 
from  2 to 3 times larger than the radius at which most of 
the bound gas sits in the FI runs after the
first pericenter passage, hence after bar formation and inflow. This is
because the remaining bound gas is that material which flows
to the center of the bar.

One interesting feature of cooling runs is that sometimes re-accretion of
gas is observed. This is clearly shown by the sequence in Figure 5
(compare the location of the stars and the gas at the two different times)
and also shows up in some of the curves of the mass loss (top panel
in Figure 13). 
This happens when gas is stripped from the disk near
pericenter but does not have enough momentum to escape the galaxy. As
the galaxy moves towards apocenter, where relative velocities are lower,
the gas falls back onto the disk.
This is not observed in adiabatic runs because gas that has been removed from 
the disk quickly becomes hotter than the virial temperature of the
dwarf halo.

%We now need to assess quantitatively the difference between
%stripping by tides alone and that by the combined action of tides and
%ram pressure. Tidal stripping in dwarf galaxies modeled similarly as 
%done here has already been studied by Mayer et al. (2001a,b) where it was 
%found  that galaxies were always retaining $20-50\%$ of their original gas
%content. However, in order to be able to draw a quantitative comparison 
%we have rerun (see Table 1)
%a few of simulations without including the gaseous halo of the primary
%(note that the simulations of Mayer et al. (2001) were using a static
%isothermal potential as opposed to the live NFW halo used here, however we
%took care that the enclosed halo mass within the pericenter of the orbit of
%the dwarfs is comparable, this being the effective perturbing mass).

We repeated several of these simulations without a gaseous halo to isolate
the efficiency of gas removal by tides alone.
The amount of gas stripped is a factor of 3 to 6 less than the simulations
which include ram pressure stripping, depending on the radiative 
cooling (Figure 13).
%We have run these simulations only with adiabatic gas to limit the 
%computational burden. 
When compared with the WT runs where only ram
pressure is included, tides strip from similar to appreciably lower
amounts of mass (lower panel of Figure 13).
The relative increase of stripping when ram pressure is added
to tides is maximum for model V40c20, that has the 
steepest rise of the dark matter density in the center (see Figure 1). This
reflects the different physical scalings of ram pressure and tidal forces.
In these highly concentrated halos the response to tidal shocks is nearly
adiabatic near the center (Taffoni et al. 2003; 
Kazantzidis et al. 2004), hence stripping is expected to be minimal
inside $R_{v=v_{peak}}$ . However, this is applicable only to stripping
induced by tidal shocks; the timescale on which istantaneous
ram pressure acts is instead much smaller than that of tidal shocks,
$< 10^7$ years (we follow Mori \& Burkert 2000 
for parameters of e.g. run Rc20V60RC)
against $\sim 10^8$ years ($\sim R_{peri}/V_{peri}$). Such timescale is 
also smaller than the local gas orbital time even in a highly 
concentrated halo ($> 10^7$ years) hence the response of the galaxy to 
ram pressure is impulsive instead of adiabatic. On the other end, the other 
models, having low
concentration haloes, respond more impulsively 
to tides in the first place, so that the increase in 
stripping efficiency is not so dramatic.

\begin{figure*}
\hskip 0.2truecm
\epsfxsize=5.5truecm \epsfbox{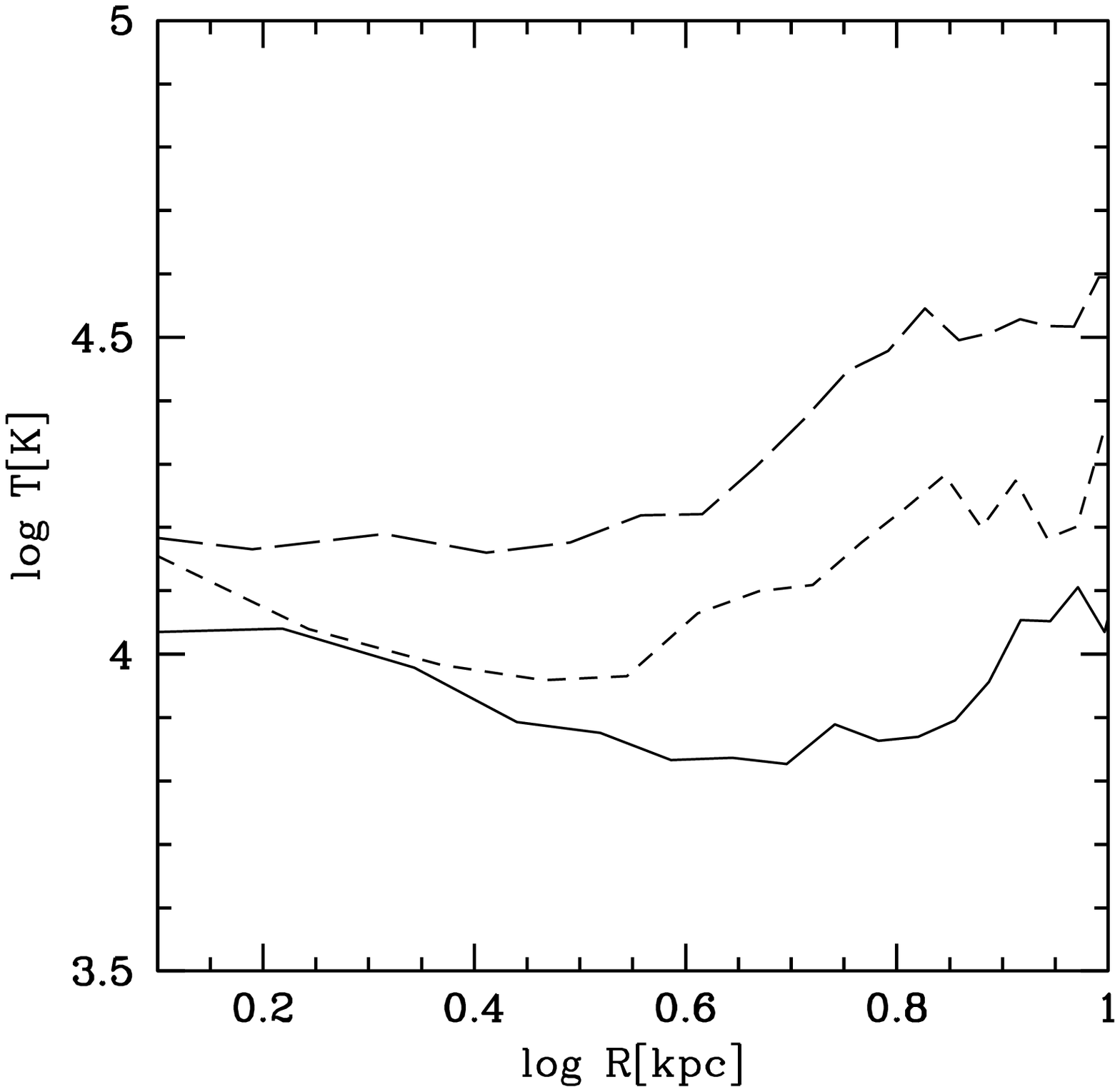}
\hskip 0.5truecm
\epsfxsize=5.5truecm
\epsfbox{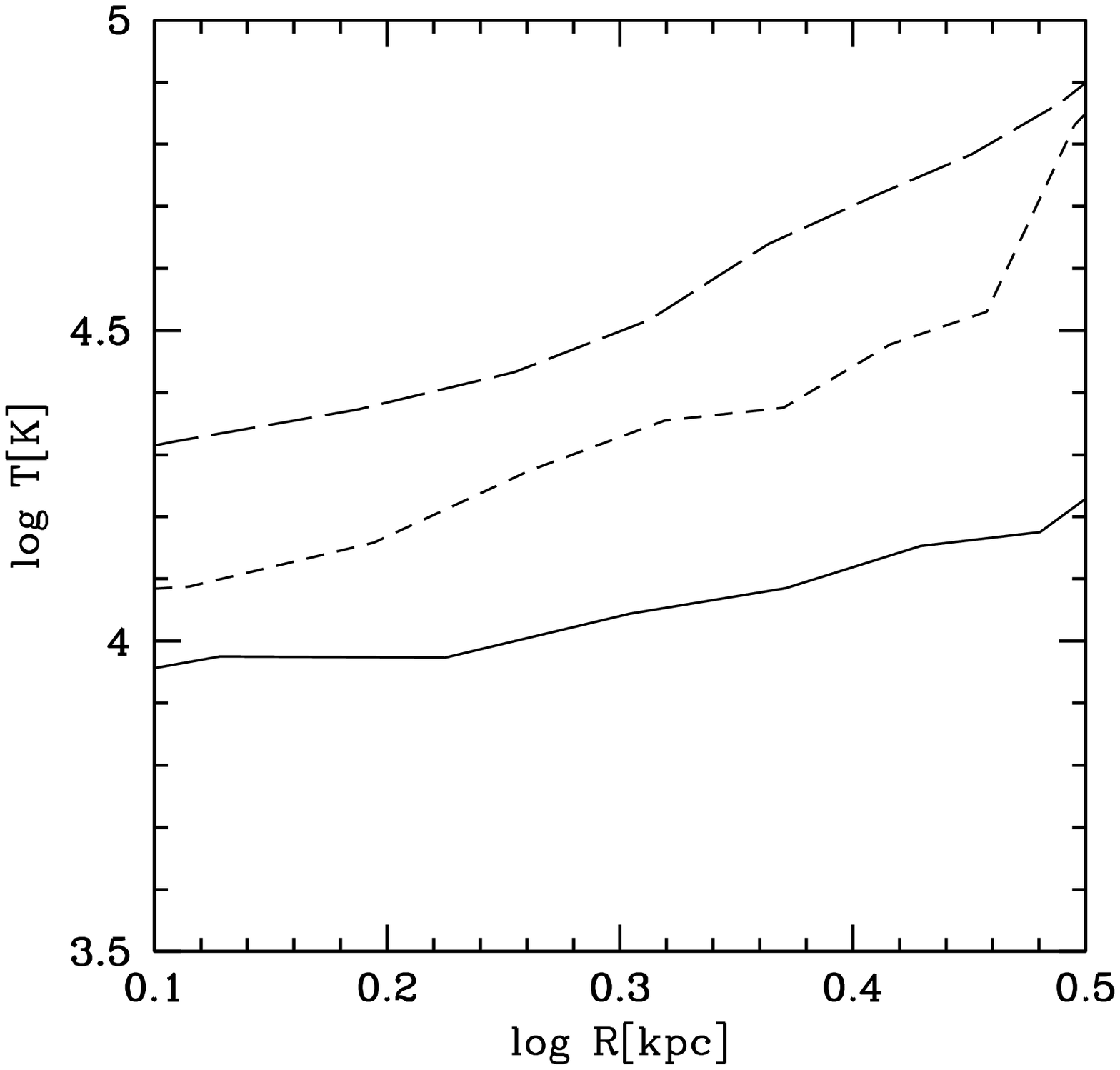} 
\epsfxsize=5.5truecm
\epsfbox{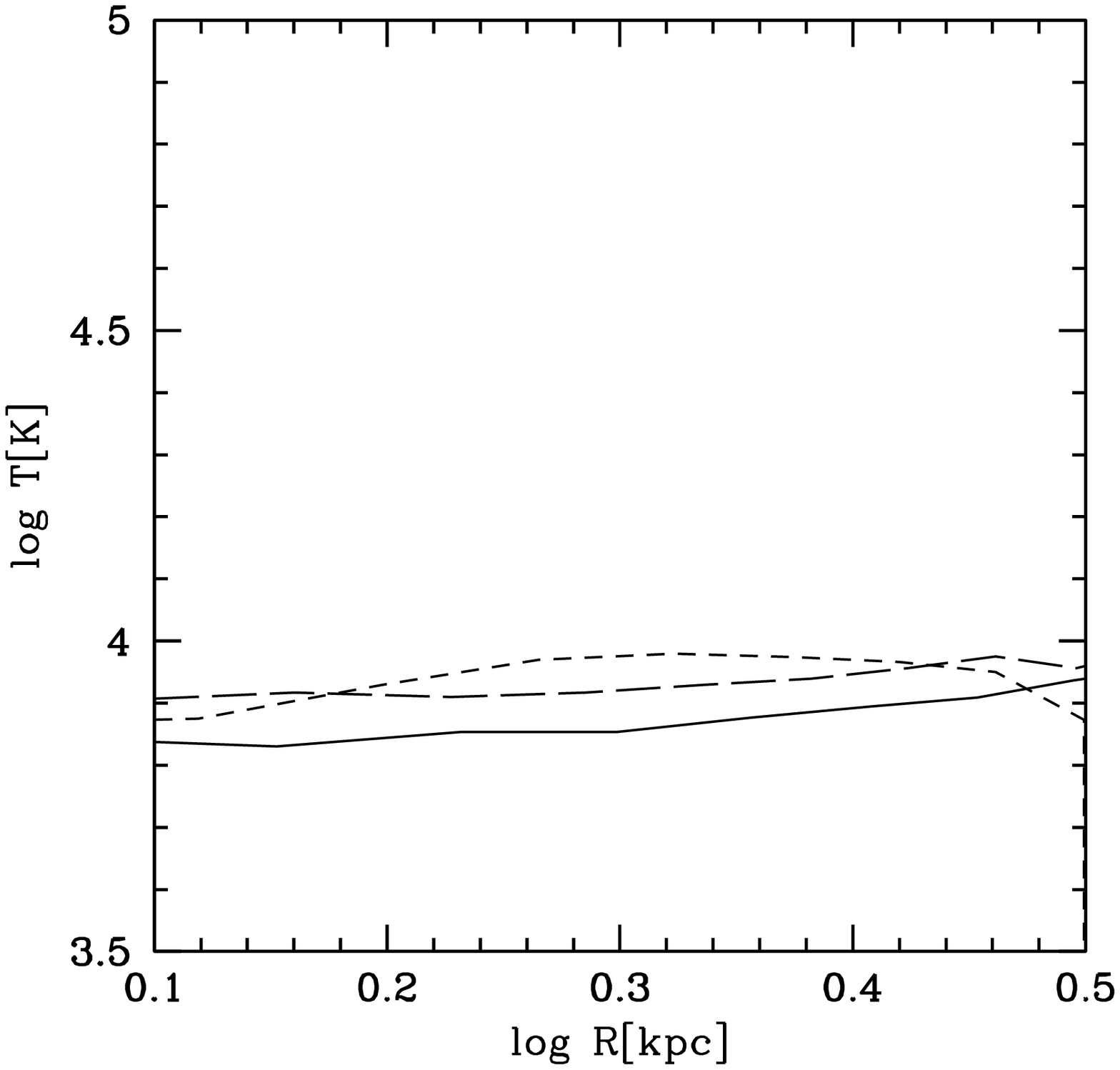}
\caption{Evolution of temperature profiles of the gas in WT runs. 
Profiles are azimuthally averaged in a cylinder with height twice that 
of the stellar disk ($R$ is the cylindrical radius).
The solid line is for $t=0.02$ Gyr, the
short-dashed line for $t=0.1$ Gyr and the long-dashed line for $t=0.3$ Gyr.
From left to right, run T1c4b, T1c20 and T2c20 are shown}.
\end{figure*}

In these simulations the peak velocity of the rotation curve of the
satellites, $V_{peak}$, drops by $20-50\%$ after several orbits (Figure  16)
while in cosmological simulations it often decreases by more than
a factor of 2 over several Gyr (Kravtsov et al. 2004, but see Ghigna 
et al. 1998). Such difference may be due to the limited number of orbital
configurations here explored (some of the satellites in Kravtsov et al. (2004)
have pericenters even smaller than 30 kpc) and/or to the fact that 
cosmological runs do not include baryons and do not have
enough force and mass resolution to follow the complex disk
evolution, especially bar formation and the resulting gas inflows that
can produce a remarkable increase in the central density 
(our force resolution is 5 times
better than the best published cosmological simulations with baryons).
These issues will be investigated systematically in
a forthcoming paper (Kazantzidis, Mayer et al., in preparation).

In summary, the final gas fractions 
drops to $0-50$ \% of the initial gas content for galaxies on 
orbits with pericenter of $50$ kpc,  
with $10\%$ being a typical fraction, and to as low as $0-6$ \%  of the
initial values for galaxies on an orbit with pericenter of $30$ kpc
(see Table 1).
In the runs that include the effect of the cosmic UV background even when
there is gas left inside the dwarf this is ionized, and the observational
upper limit on the mass of neutral hydrogen in dSphs, $M_{HI} < 0.01$
$M_{s}$, where $M_s$ is the stellar mass, is fully satisfied. 
The gas will be able to
recombine later on as the cosmic background fades away, but the local 
ultraviolet radiation flux from the Milky Way or M31 might be enough to keep
it ionized even at low redshift (Maschenko, Carignan \& Bouchard 2004).
The current upper limits on the total amount of gas in dSphs including the 
ionized component, are much higher, $M_g < 0.1 M_s$ (Gallagher et al. 2003)
, and are satisfied in the final states of the dwarfs for the majority of 
our runs. 
Hence a combination of ram pressure and tidal
stripping succeeds in explaining how dwarf spheroidals have little or no
neutral hydrogen today.
On any given orbit, dwarf models with higher $V_{peak}$ are
stripped less severely and, in particular, our most massive model, V60c4,  
($V_{peak} = 62 $ km/s), 
always retains a substantial amount of gas. This is consistent with the fact
that  bright dwarf elliptical
(dEs) satellites of M31, like NGC205, whose central velocity dispersion, 
$\sigma = 30-45$ km/s, is comparable to that of the evolved states
of model V60c4, do have some amount of neutral gas (Mateo 1998). 

\begin{center}
\begin{figure*}
\hskip 0.3truecm
\epsfxsize=4truecm 
\epsfbox{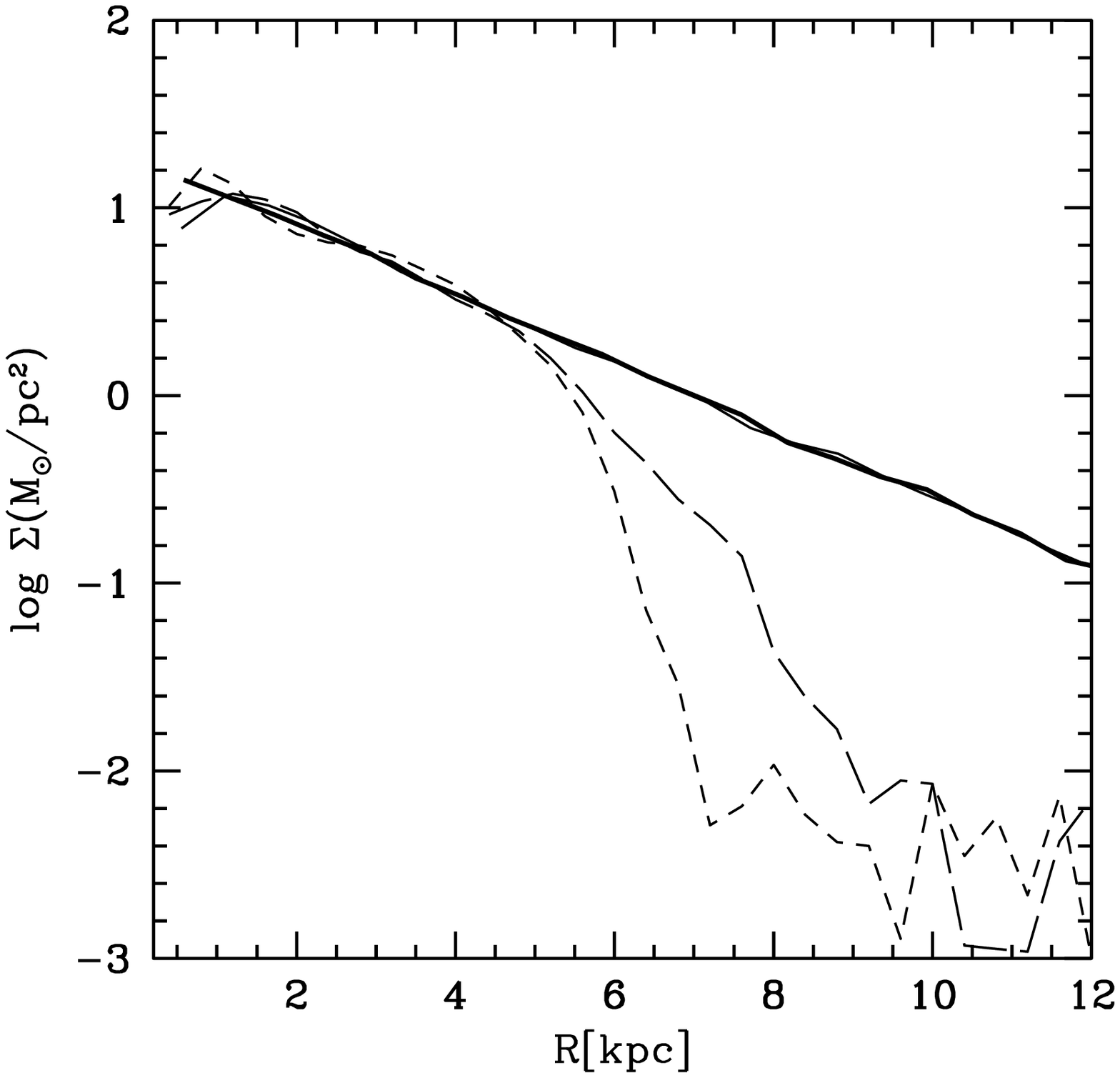}
\epsfxsize=4truecm 
\epsfbox{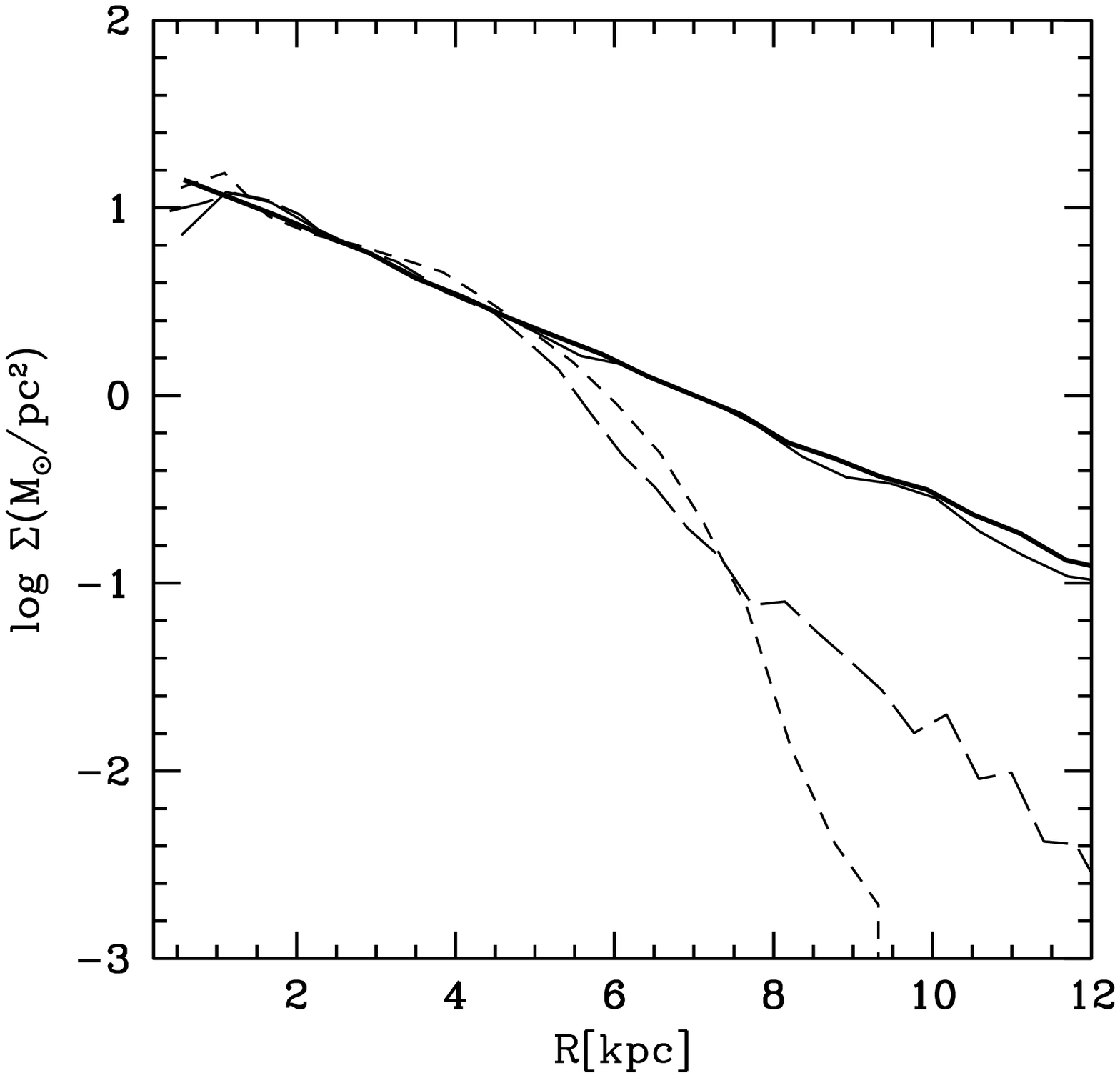}
\epsfxsize=4truecm
\epsfbox{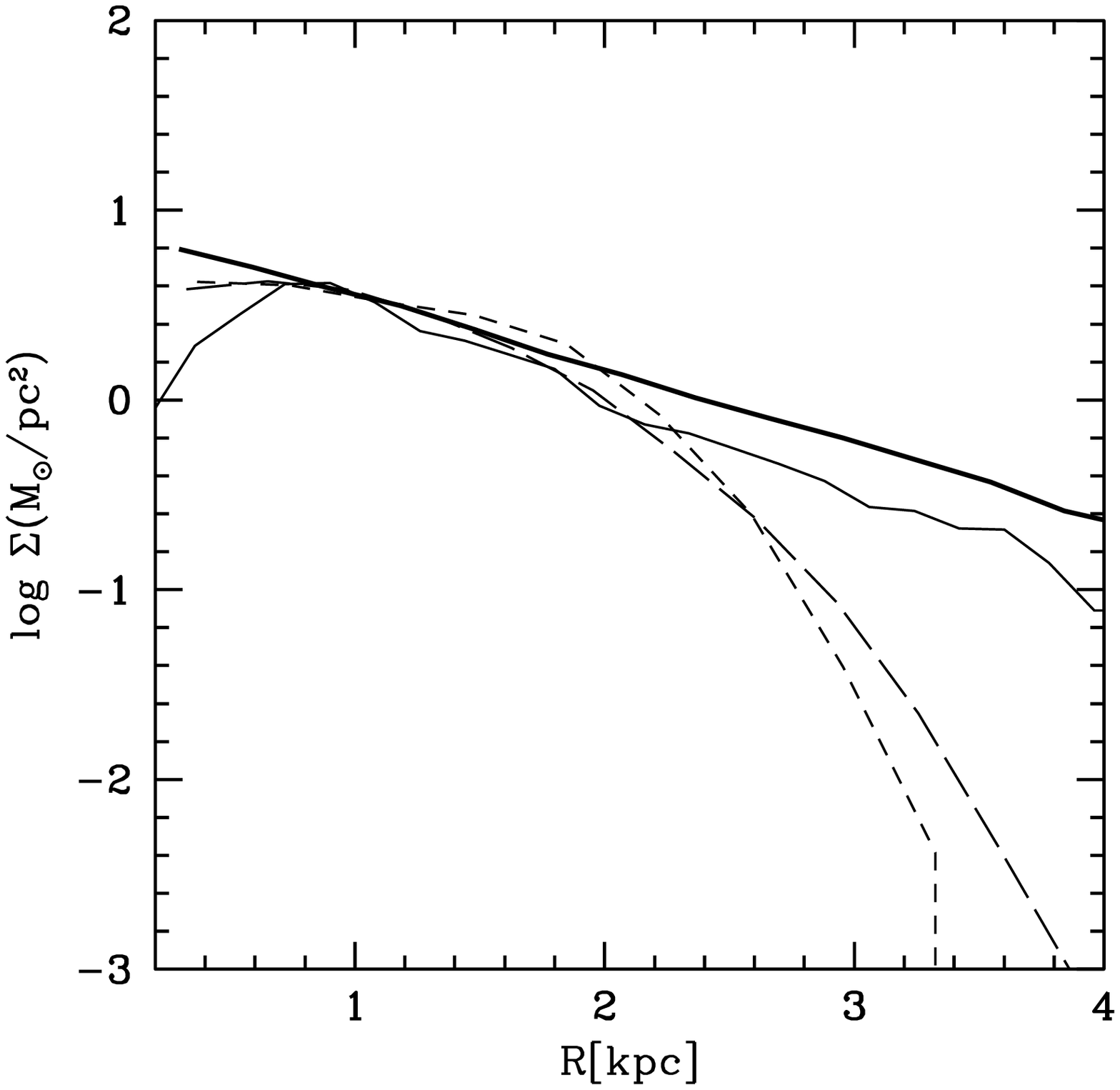} 
\epsfxsize=4truecm
\epsfbox{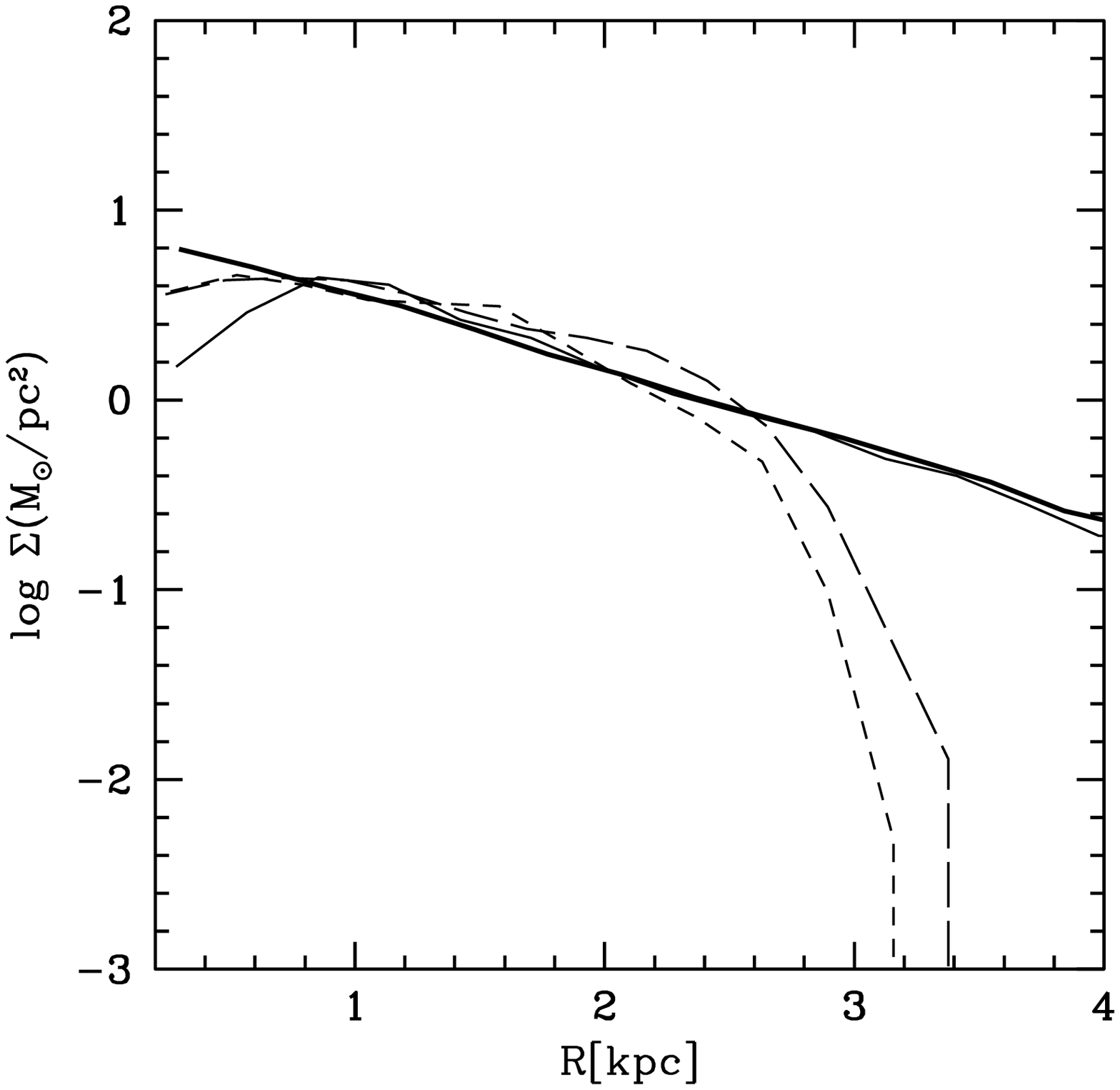}
\caption{Evolution of density profiles of the gas in WT runs. 
Profiles are azimuthally averaged in a cylinder with height equal to that of 
the stellar disk ($R$ is the cylindrical radius).
The thick solid like is for $t=0$, the thin solid line for $t=0.02$ Gyr, the
short-dashed line for $t=0.1$ Gyr and the long-dashed line for $t=0.3$ Gyr.
From left to right, runs T1c4b, T3c4b, T1c20 and T2c20 are shown}.
\end{figure*}
\end{center}

When neutral gas remains in the dwarfs, most of it  
will be rapidly turned into 
stars as it reaches very high densities due to the
effect of the bar, and bursts localized near the center of the dwarf 
are expected at pericenter passages.
as described in Mayer et al. (2001a). Hence we predict that the higher 
the mass of the dwarf (not its luminosity) the more extended its star 
formation history should be.
%However, a similar trend is not expected with luminosity since dwarfs with
%initially equal stellar mass, and thus luminosity, can be affected very
%differently by these environmental mechnaisms depending on the mass
%and concentration of ther halo (compare runs with model V28c4 and those with
%model V40c20 in Table 1). In addition, 
For dwarfs that start out with similar masses, our simulations suggest 
that the outcome will vary depending on their orbit. 
%as we just saw, therefore one should not expect clear trends between gas
%content/star formation history and the luminosity of the dwarf. Therefore
%the fact that such clear trends do not exist in LG dSphs is not surprising
%within this scenario.  What 
A signature in the star formation history of the dwarfs is expected
at the epoch when it underwent the first close approach within the
Milky Way. This can be either a burst or a marked depression 
of the star formation depending on whether just a fraction or most of the
gas is removed from the dwarf galaxy.
%Our results indicate that the 
%second situation is more likely to have happened  for dwarfs that fell in
%early, at $z > 1$, and that today are found closer to their parent
%galaxy. A similar trend has been found also
%in cosmological simulations of galaxy formation (Governato, Mayer et al. 2004;
%Governato, Mayer et al., in preparation).
%An earlier truncation of the star formation for the closest Milky Way
%satellites, Draco, Ursa Minor and Sculptor, is indeed observed (Van der Bergh
%1993).

\subsection{Kelvin-Helmholtz instabilities and turbulent stripping}

Gas that survives istantaneous ram pressure can be stripped by the 
Kelvin-Helmholtz
(KH) instability developing at the interface between the gas of the 
dwarf galaxy and the ISM (Nulsen 1982; Murray et al. 1993, Quilis et al. 2000).
This is also known as turbulent stripping since
the gas reaches a turbulent state as the instability develops.
In order to develop KH instabilities require a sharp boundary to exist between 
the two fluids. 
Even at the high resolution adopted here SPH is hardly capable of resolving
KH instabilities mostly because the artificial viscosity and the fact that
densities are smoothed in SPH both contribute to ``blur'' any  sharp 
interface.
An approximate ``effective''local kinematical viscosity can be calculated as 
$\nu=\alpha c_s h/8$, where $\alpha$ is the coefficient
of the linear term in the artificial viscosity, $c_s$ is the sound speed
and $h$ is the SPH smoothing length (Lufkin et al. 2004).
%This is an approximate estimate of the
%mean viscosity in the simulated fluid; for particles participating to the 
%bow shock the quadratic $\Beta$ term of the Monaghan standard formulation
%adds to the artificial viscosity. On the other end the Balsara correction 
%reduces the viscosity by a factor of $5-10$ since we are simulating a 
%shearing flow,  hence considering only the viscosity contributed by
%the linear term should yield a reasonable estimate.
The Reynolds number is then $Re \sim Lv_{gal}/\nu$,
where $L$ is the typical size of the object moving in an external medium
(the disk of the dwarf here) and $v_{gal}$ is the speed at which it is moving;
this number is strictly a function of position 
(since $\nu$ is a function of
position) and is in the range $20-100$ in our runs
(the highest Reynolds numbers occur in the highest resolution WT runs).
This is of course much lower than the Reynolds
numbers typical of turbulent flows (up to $10^4$). Aside from numerical
issues that might hamper the development of KH instabilities in the simulations
we can ask whether turbulent stripping is likely to be important for dwarf
galaxies moving in a very diffuse halo as those that we are modeling here.
Mori \& Burkert (2000) found that spherical dwarf galaxies moving at a few
thousand km/s in a dense cluster hot halo will be stripped on a timescale
significantly smaller than the crossing time of the galaxy in the cluster. 
Following Mori \& Burkert (2000) we use the following equation to obtain an
estimate of the expected characteristic timescale for stripping by
KH instabilities including the stabilizing effect of gravity (which is mainly
provided by the dark halo potential in our models);

\begin{eqnarray}
\tau_{\rm KH}&=&\frac{F M_0}{\dot{M}_{\rm KH}}, \\
&=&2.19\times10^9
 \left(\frac{F          }{0.1                     }
                                        \right)
 \left(\frac{M_0        }{10^{ 9} M_\odot         }
                                        \right)^\frac{1}{7}
 \left(\frac{n_{h}}{10^{-4}{\rm cm           }^{-3}}
                                        \right)^{-1}
                                        \nonumber \\
&& \makebox[7em]{} \times
 \left(\frac{v_{\rm gal}}{10^{ 3}{\rm km~s}^{-1}}
                \right)^{-1} ~{\rm yr}. \label{taukh}
\end{eqnarray}

The equation assumes spherical symmetry, which is a reasonable assumption
since our galaxies are dominated by a spherical dark matter halo.
In the equation $M_0$ is the CDM halo mass within the radius $
R_{strip}$ down to which galaxies are stripped in the WT runs
(see Table 2, these simulations offer a cleaner test case since they include 
only ram pressure), $F=M_{bar}/M_{0}$ is the mass ratio between the baryons
and the CDM halo within the same radius, and $\dot{M}_{\rm KH}=\pi R_{strip}^2 \rho_{h} v_{\rm gal}$ is the mass loss rate from the galaxy through
KH instabilities ($\rho_h$ is the hot halo density and $v_{gal}$ is the
velocity of the galaxy through the hot halo). For the gas densities
($\sim 10^{-4}$ atoms cm$^{-3}$) and velocities occurring at pericenter
 the timescale for KH stripping 
is, respectively, 4 Gyr for model V40c20 and 10 Gyr for model V60c4,
these being the model galaxies in which instantaneous ram pressure is not 
capable of removing all the gas.
For the gas densities and velocities where the satellites spend most
of their time, namely at distances larger than 100 kpc, the KH stripping 
timescale is at least several Hubble times.

%The timescale is ten times shorter, 
%and therefore KH stripping would be non-negligible, for 
%the model with the shallowest potential well, V28c4, but in that case 
%instantaneous ram pressure is sufficient at removing entirely the gas 
%anyway (see Table 2).

\begin{figure*}
\hskip 1.1truecm
\epsfxsize=6truecm \epsfbox{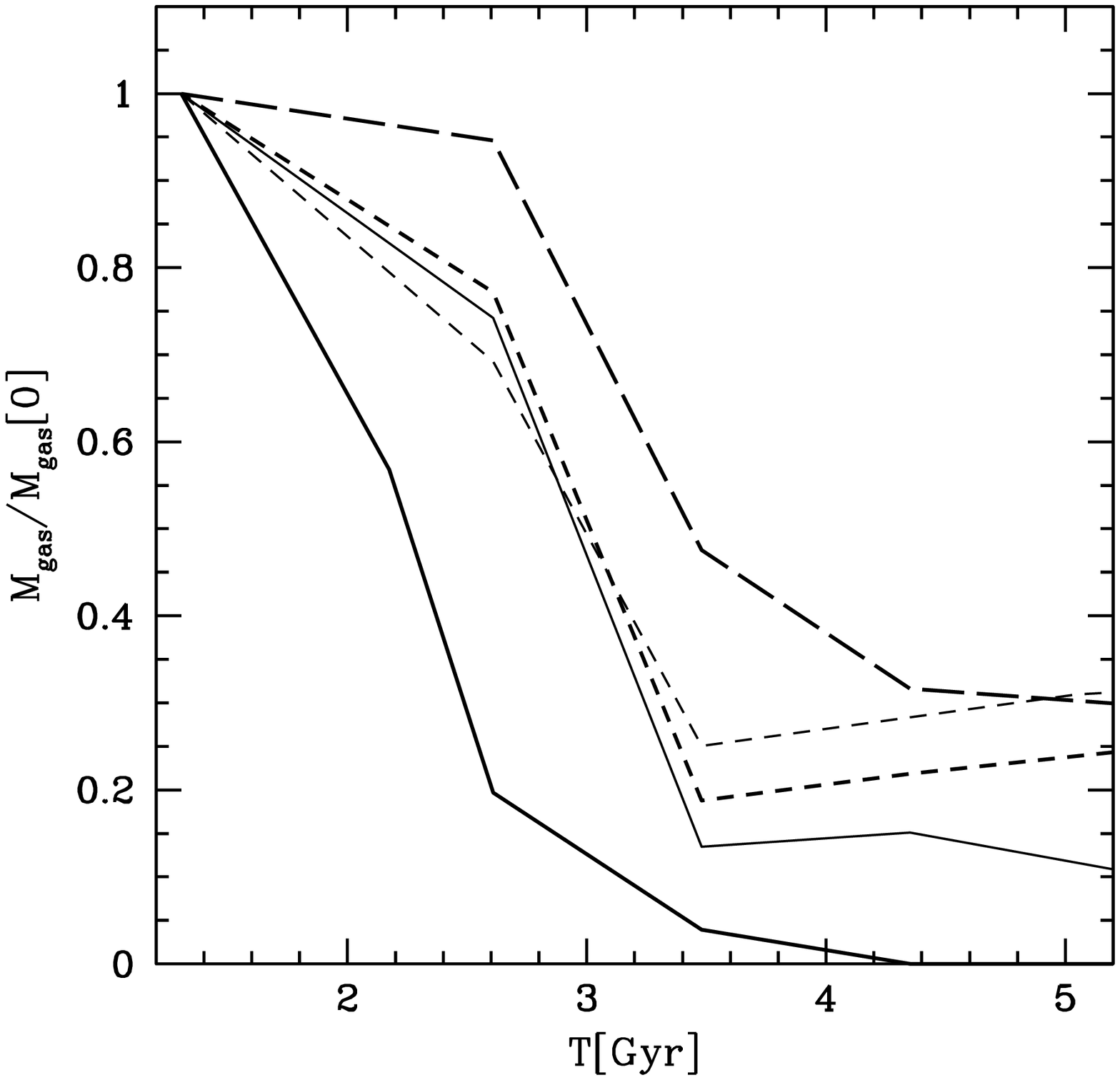}
\end{figure*}
\begin{figure*}
\hskip 1.1truecm
\epsfxsize=6truecm
\epsfbox{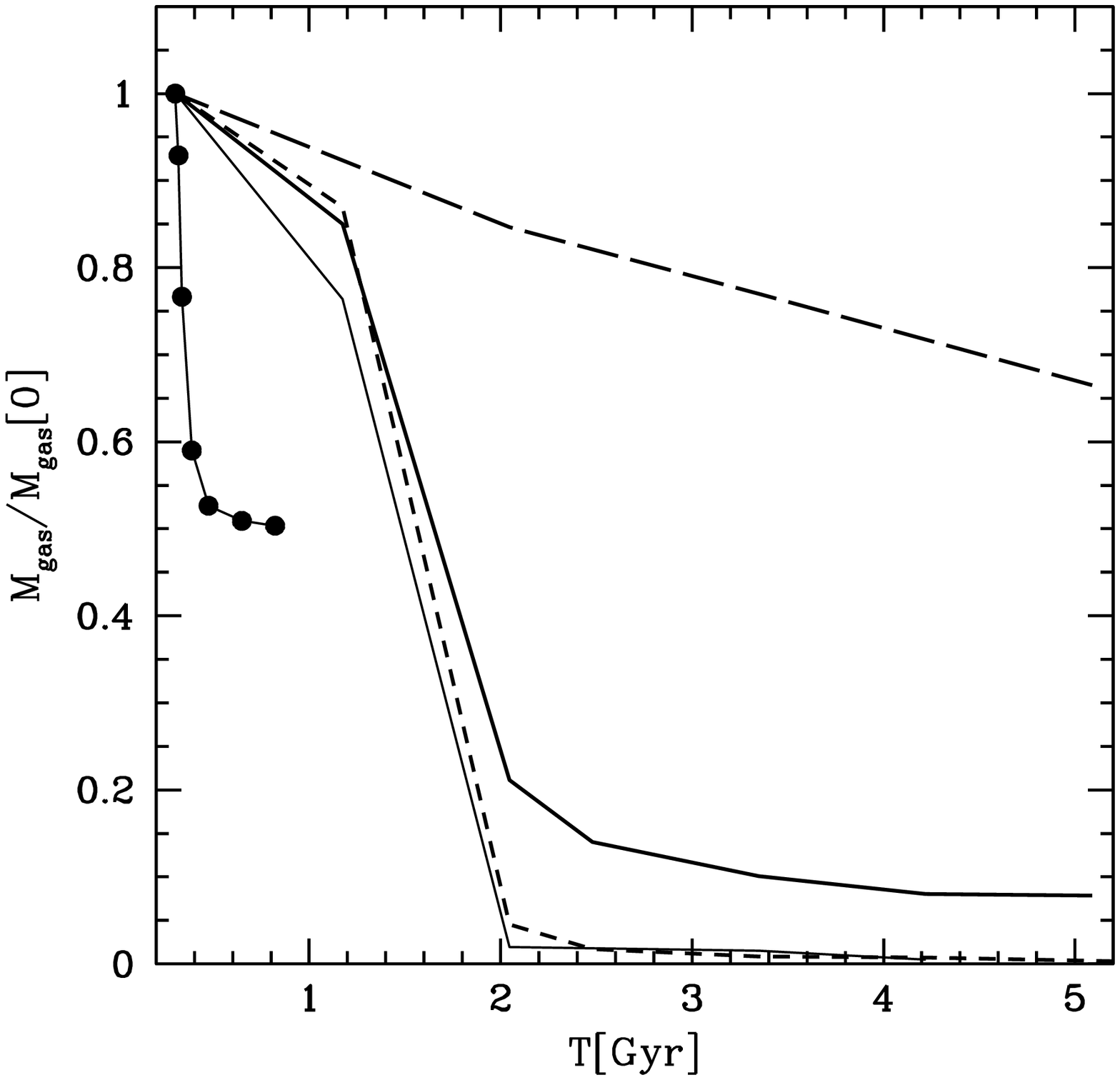}
\caption{Time evolution of the bound gas mass for FI runs using model 
V60c4, V28c4 (top) and runs using model V40c20 (bottom). Top: run Rc4aAD
(thin solid line), Rc4aRCUV (thin short-dashed line), Rc4bTD (long-dashed line),Rc4bAD (thick solid line), Rc4bRC (thick short-dashed line).  
Bottom: run Rc20TD (long dashed line), Rc20AD (thin solid line), Rc20RC
(thick solid line line), Rc20RCUV (short dashed line). The line with dots
shows the mass loss rate (due to ram pressure only) for one of the WT runs, 
T1c20  (see Table 2 -- the curve stops at less than 1 Gyr since WT
runs were run for a short timescale since stripping 
saturates very early on.}
\end{figure*}

Shear viscosity in a laminar flow ($Re < 30$) can also produce a slow stripping
of the gas with timescales and mass loss rates comparable to those of
KH stripping in the turbulent regime (Nulsen 1982). Physical viscosity 
is not included in our treatment of SPH and therefore this viscous 
stripping mode is missing. Spurious viscous stripping from the artificial 
viscosity term is limited by our reasonably high resolution and the Balsara 
correction.
The characteristic viscous time is $\tau= L^2/\nu$ is $\sim$ 10 Gyr, 
which means viscosity effects are negligible given that the important
timescales, those related to ram pressure and tidal shocks, are much smaller 
(between $10^7$ and $10^8$ years). 
Tittley, Pearce \& Couchman (2001) have shown that SPH can alter the magnitude 
and scaling with density of hydrodynamical drags because of the way 
pressure forces among neighboring cold and hot background particles are 
calculated (in essence the cross section of an object moving in a hot 
medium can be artificially enhanced or decreased). However, the effect is 
particularly severe, enhancing the drag over the 
expected value, only in the subsonic regime, whereas our galaxies all move 
at slightly supersonic velocities. 

In summary, our simulations have enough resolution to overcome artificial
viscous stripping but cannot model correctly turbulent stripping yet.
However, although SPH is often blamed for its inability to model 
hydrodynamical 
turbulence we recall that published grid simulations are of similar 
quality when it comes to the ability to model turbulence -- 
Marcolini et al. (2003) find Reynolds numbers similar to what we find.
%A broad code comparison effort currently on the way (Mayer et al., in
%preparation) is finding however that SPH can achieve results similar to 
%adaptive mesh refinement codes and model turbulent stripping with comaparable 
%accuracy once the resolution is increased by another order of magnitude 
%relative to that used here and special care is used in the preparation of the
%initial conditions.

\subsection{Dark and baryonic contents of the remnants: can we explain Draco?}

In FI runs the stellar disk is stripped and transformed into a spheroidal
system by means of tidal heating and bar-buckling instabilities as 
described in Mayer et al. 2001a,b. The ratio $M_{dark}/M_{bar}$ 
between dark matter mass and baryonic mass in the final objects varies between 
$3$ and $12$, the highest values occurring in runs with model V40c20.
In remnants with the highest dark matter contents or highest central 
concentration of gas the buckling instability is weaker, with the result that 
the bar-like shape is preserved. 
The trend with gas concentration is consistent 
with the findings of Debattista, Mayer et al. (in preparation) 
for isolated bar-unstable 
galaxies with gas. However, in reality the central gas would form stars
and its distribution will become less concentrated. 
%We are investigating
%further these issues in a companion paper in which we focus on the structure
%and kinematics of the stellar remnants (Mayer et al., in preparation). 
The final
central $v/\sigma$ of the stars is below 0.5, thus compatible with
those of dSphs, and the central final velocity dispersions are in the range
$10-30$ km/s. 

We can ask whether the remnants of model V40c20, can reproduce the overall 
properties
of Draco, one of the most extreme dSphs in terms of low luminosity and
high $M/L$.
%Let us begin with the final baryonic content in the runs employing such
%model (see Table 1). 
In adiabatic runs the gas is completely stripped
after the first pericenter passage. The observed truncation of the star
formation history in Draco could be thus interpreted as the result of its
infall into the Milky Way halo. However, in runs with cooling some
gas is retained. Assuming a
star formation rate comparable to that deduced for Draco and Ursa Minor
by Hernandez et al. (2000), 
$100-200 M_{\odot}$/Myr, we obtain that the
remaining gas in  run Rc20RC will be turned into stars in about 5 Gyr,
leaving a dwarf completely devoid of gas. 
But this implies a fairly extended star formation
history, contrary to what observed for Draco, and
the amount of gas that needs to be turned into stars
is non-negligible, $\sim 6 \times 10^6 M_{\odot}$; if we sum such mass
to the remaining bound stellar mass we get 
$5 \times 10^7 L_{\odot}$. For a mass-to-light ratio $\sim 4$, as typical
of old stellar populations we would get a luminosity $\sim 10^7 M_{\odot}$,
i.e. 50 times higher than the V band luminosity of Draco (Odenkirchen et
al. 2001). 
%Note that in such $c=20$ halos tidal
%stripping of stars is minimal (see Mayer et al. 2002), this being 
%consistent with the fact that tidal tails have not been detected in
%Draco down to $32$ mag arcsec$^{-2}$ (Odenkirchen et al. 2001).
%This luminosity is closer to that of the much brighter dSph Fornax.
Since the largest fraction of the final bayonic mass is contributed by
the initial stellar mass in the model (tidal stripping is moderate in
highly concentrated haloes) an excessive final luminoisty would result
also in adiabatic runs.
As for the dark matter content one would obtain
$M/L =\sim 50$, as high as typical estimates for $M/L$ in Ursa Minor and 
comparable to the lower limits for $M/L$ in Draco (Mateo 1998; 
Wilkinson et al. 2004). The final projected central velocity
dispersion, $\sim 15$ km/s, is consistent with the peak velocity
dispersion of Draco (Wilkinson et al. 2004). The central 
dark matter density measured
for the stirred dwarf at the end of the simulation, $0.065 M_{\odot}/pc^3$ 
is also in rough agreement with the value estimated for Draco 
($\sim 0.1 M_{\odot}/pc^3$). Thus run  Rc20RC produces a model with 
roughly the right dark matter content but an excessive baryonic content.
Adding heating and photoionization from
a uniform UV background (run Rc20RCUV) aids gas removal
and leaves essentially no neutral gas 
after a few Gyr. In this case the star formation history, as in the adiabatic
runs, would have been essentially truncated after the first close
encounter with the Milky Way. 
Nevertheless, the total stellar mass of the dwarf would still yield a 
luminosity about an order of magnitude higher than that of Draco.

\begin{figure*}
\hskip 1truecm
\epsfxsize=6.6truecm 
\epsfbox{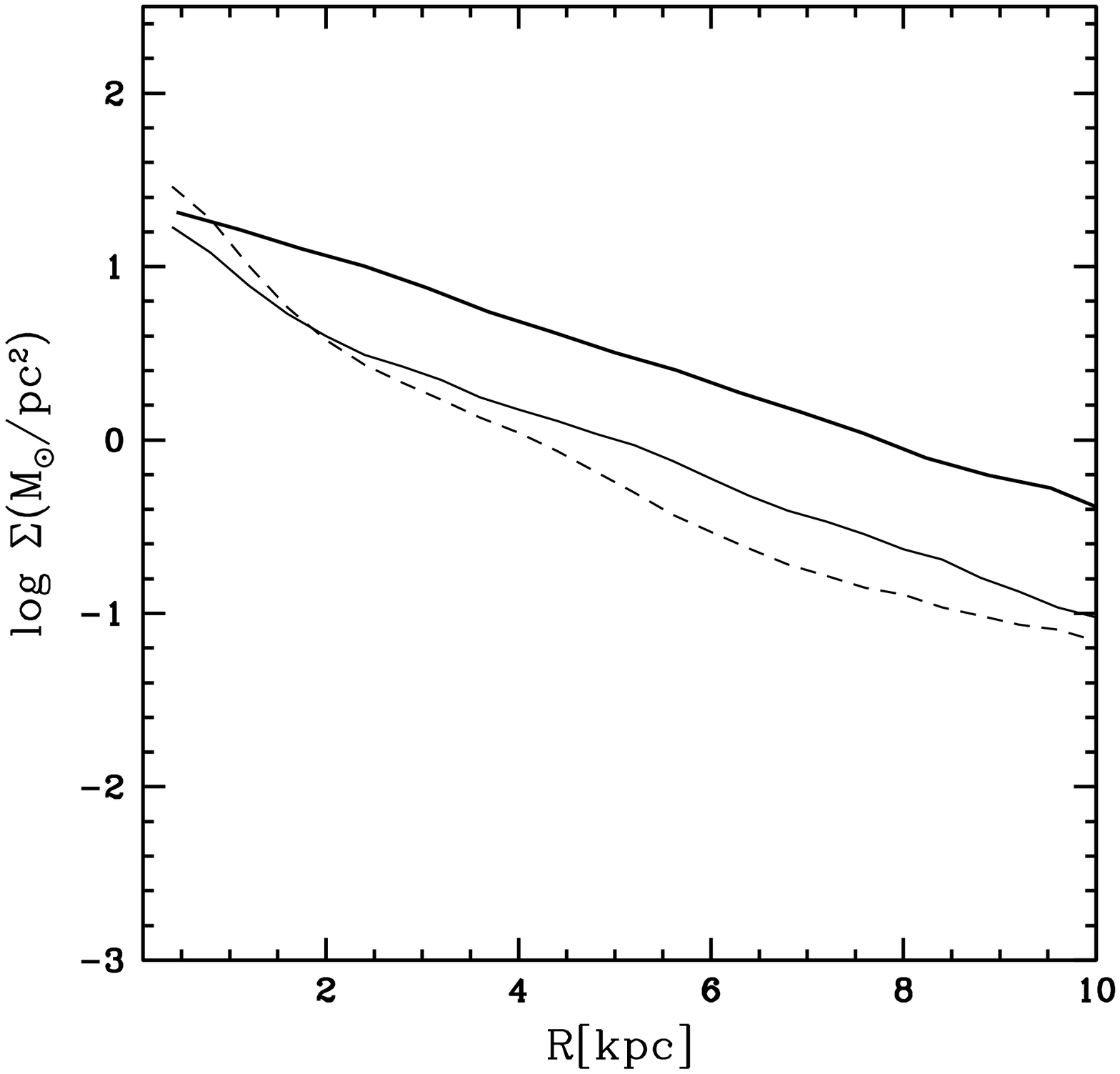}
\hskip 1truecm
\epsfxsize=7.5truecm
\epsfbox{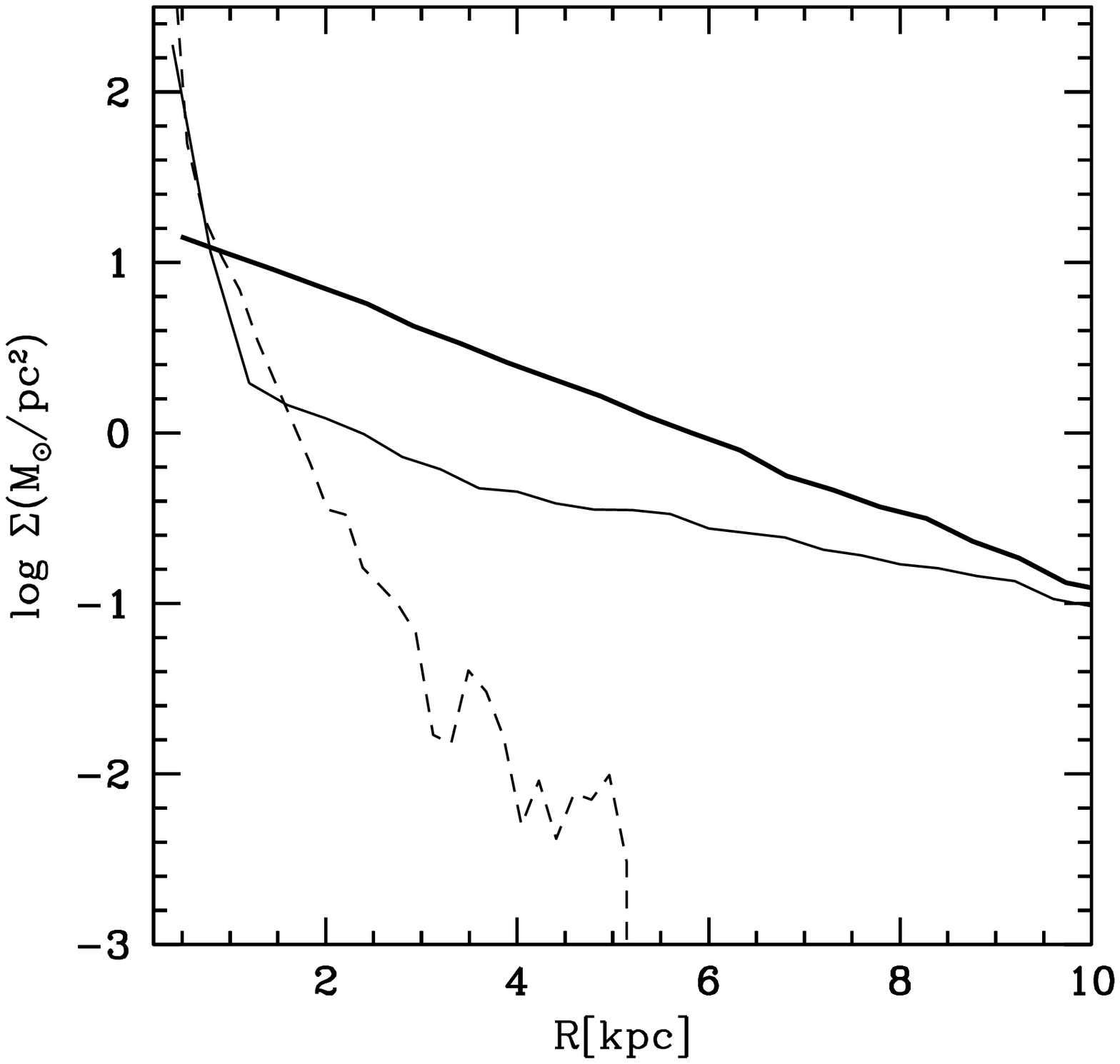}
\hskip 1truecm
\epsfxsize=7.5truecm
\epsfbox{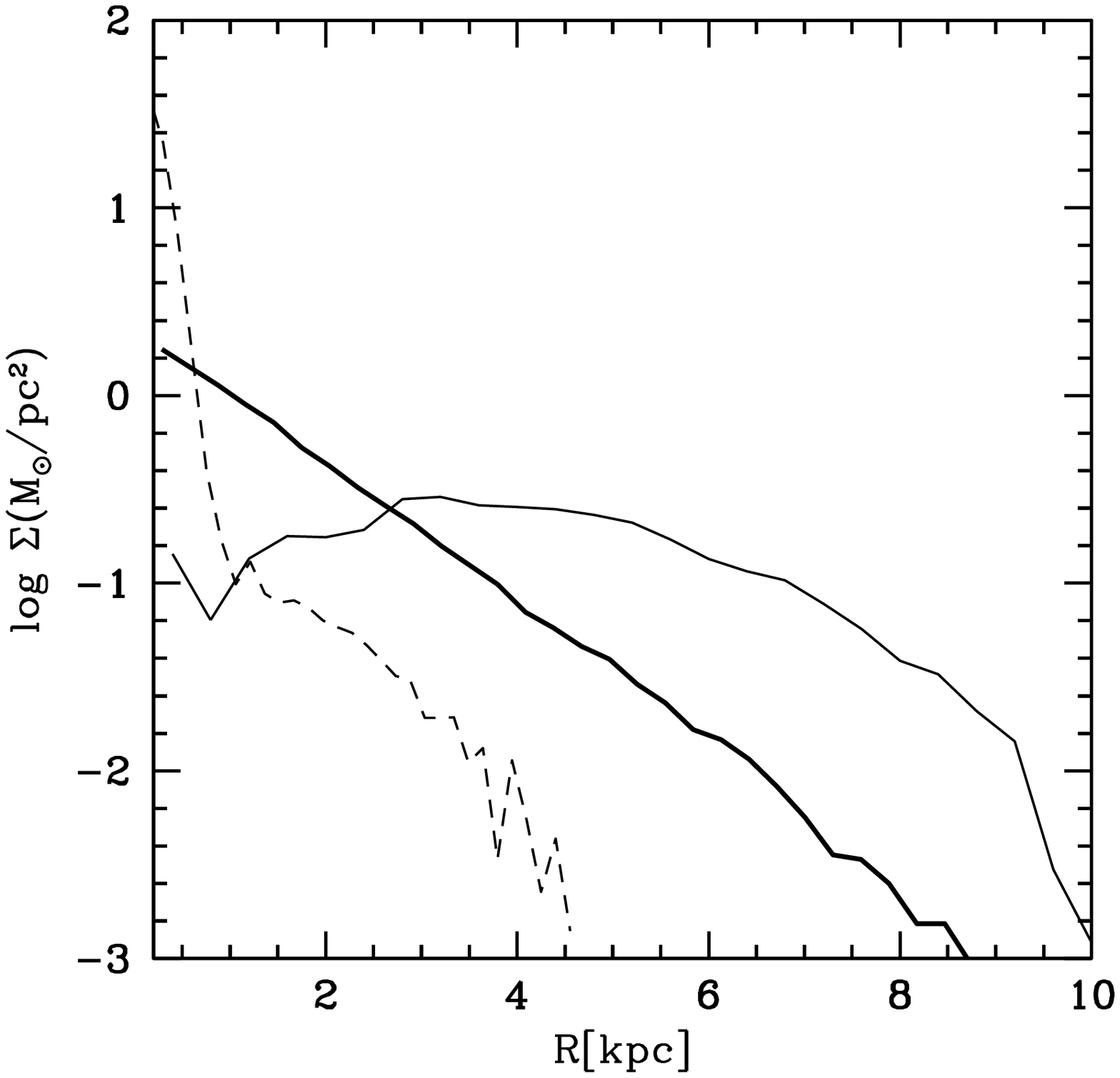}
%\epsfxsize=7truecm
%\epsfbox{v75c4UVsurfgas.ps}
\caption{Evolution of gas surface density profiles in FI runs. The thick solid 
line is used for the initial profile, the thin solid line is used for the
profile at T=3 Gyr (after first pericenter passage), and the dashed line
is used for the profile at T=6 Gyr (after second pericenter passage). 
From top left to bottom, the runs Rc4aAD, Rc4aRCUV, and Rc4bRC are shown.
The unusual shape of the profile at intermediate times in the bottom panel
reflects the fact that gas is displaced from the galaxy at this point
(see also Figure 5).}
\end{figure*}

The final luminosity depends on the initial stellar and gas content
of the model, which was chosen to be fairly representative
of the baryonic content of todays' dIrrs.
Although Draco likely had 
a halo massive enough to resist gas removal from supernovae feedback or
photoionization, the progenitors from which it assembled at very high 
redshift were certainly much smaller and were likely affected by these
mechanisms.
Hence one can imagine that the main progenitor of Draco had
a baryon fraction much lower the one we used here.
In such a scenario, however, environmental effects 
would not have played the primary role in setting the current high $M/L$
of Draco. One alternative explanation is
to imagine that Draco was a mostly gaseous disky dwarf
by the time it entered the Milky Way, with minimal star formation having 
taken place. Such a low star formation rate can be simply
the result of an extended, relatively high angular momentum disk being almost
everywhere Toomre stable and below the density threshold necessary for 
star formation to happen, as observed today in some dIrrs and LSB galaxies 
(Verde, Oh \& Jimenez 2002). To explore this possibility we have run a
simulation in which the initial model has the same structural properties
of model V40c20 except that now as much as 90\% of the disk is gaseous.
A weaker stellar bar forms given the lower stellar mass which produces
a weaker gas inflow, and a larger relative fraction of the disk mass
is stripped by ram pressure. In an adiabatic run all the gas is stripped
after two pericenter passages, leaving a gas-free remnant that 
roughly matches the luminosity of Draco after $\sim 3$ Gyr. In a run
with cooling and the cosmic UV radiation of order $10^5 M_{\odot}$ of gas
remain in the dwarf, but all this gas is ionized and no star formation
would be possible, and even in this case the final stellar mass would
be comparable to that of Draco.

%In runs in which the different dwarf models 
%are evolved on wider orbits, dwarfs usually retain enough gas
%mass after pericenter passages to undergo one or two  bursts of star 
%formation (see Mayer et al.
%2001b). The luminosities calculated from their final baryonic content 
%,(namely assuming that the gas would form stars) are comparable to 
%those of Carina (e.g. Rc4bRC) and Fornax (e.g. run Rc20bAD, Rc4AD). $M/L$
%values in remnants using model V28c4 and model V60c4 are in the range
%$8-20$, comparable to the estimates for Fornax, Carina or Leo I (Mateo 1998). 
%The strength of the bursts, assuming a Schmidt law,
%as done in Mayer et al. (2001b), is comparable to  that of the secondary
%episodes inferred for Leo I and Fornax. 

\subsection{Fate of stripped gas}

The stripped gas has a very different fate in the adiabatic and in the cooling
runs.
In the adiabatic runs it ends up quite diffuse and hot ($T > 10^5$K) 
and quickly  thermalizes and mixes with the gas in the MW halo. 
When cooling is allowed the gas along
the tails fragments into dense clouds and sheet-like structures 
pressure confined by the ambient medium. This is shown in Figure 17.
%The blobs are resolved by several hundred particles, so they
%ae well above the SPH resolution limit.
These clouds remain cold ($T < 10^5$ K) until they are
dissolved as they sink deeper in the potential well due to 
gas drag. Sinking of the gaseous tails is marginal in the adiabatic 
runs because the 
stripped gas is too diffuse to experience significant drag. Fragmentation
is suppressed in the first place in adiabatic runs because the density of the
gas along the tails is too low for the cloud to reach pressure equilibrium 
within the hot ambient medium.  In the simulations with UV heating, clouds are 
also present, although they are slightly more diffuse and hotter 
(see Figure 17).

\begin{figure*}
\hskip 0.4truecm
\epsfxsize=4.5truecm \epsfbox{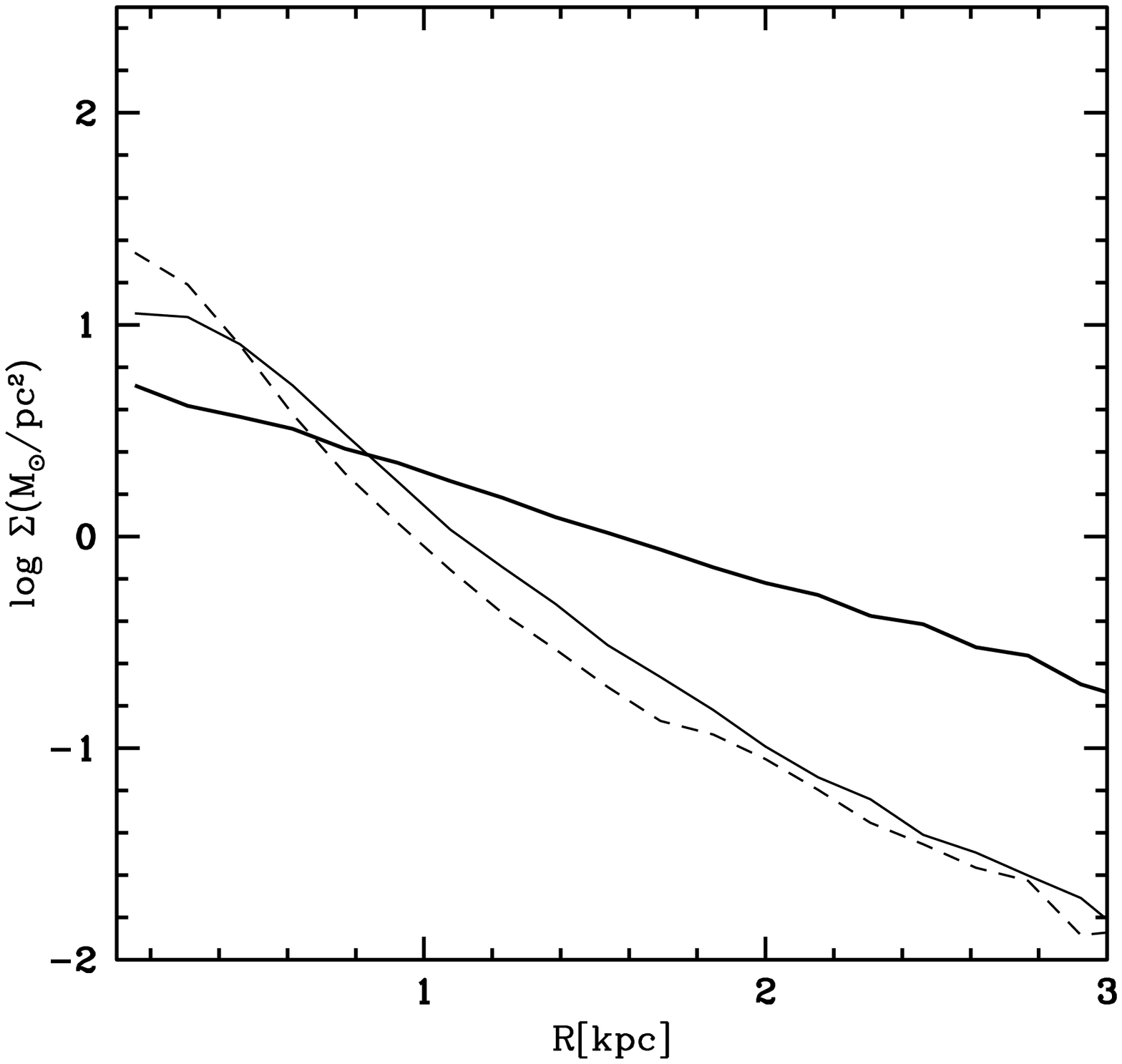}
\epsfxsize=4.5truecm
\hskip 0.4truecm
\epsfbox{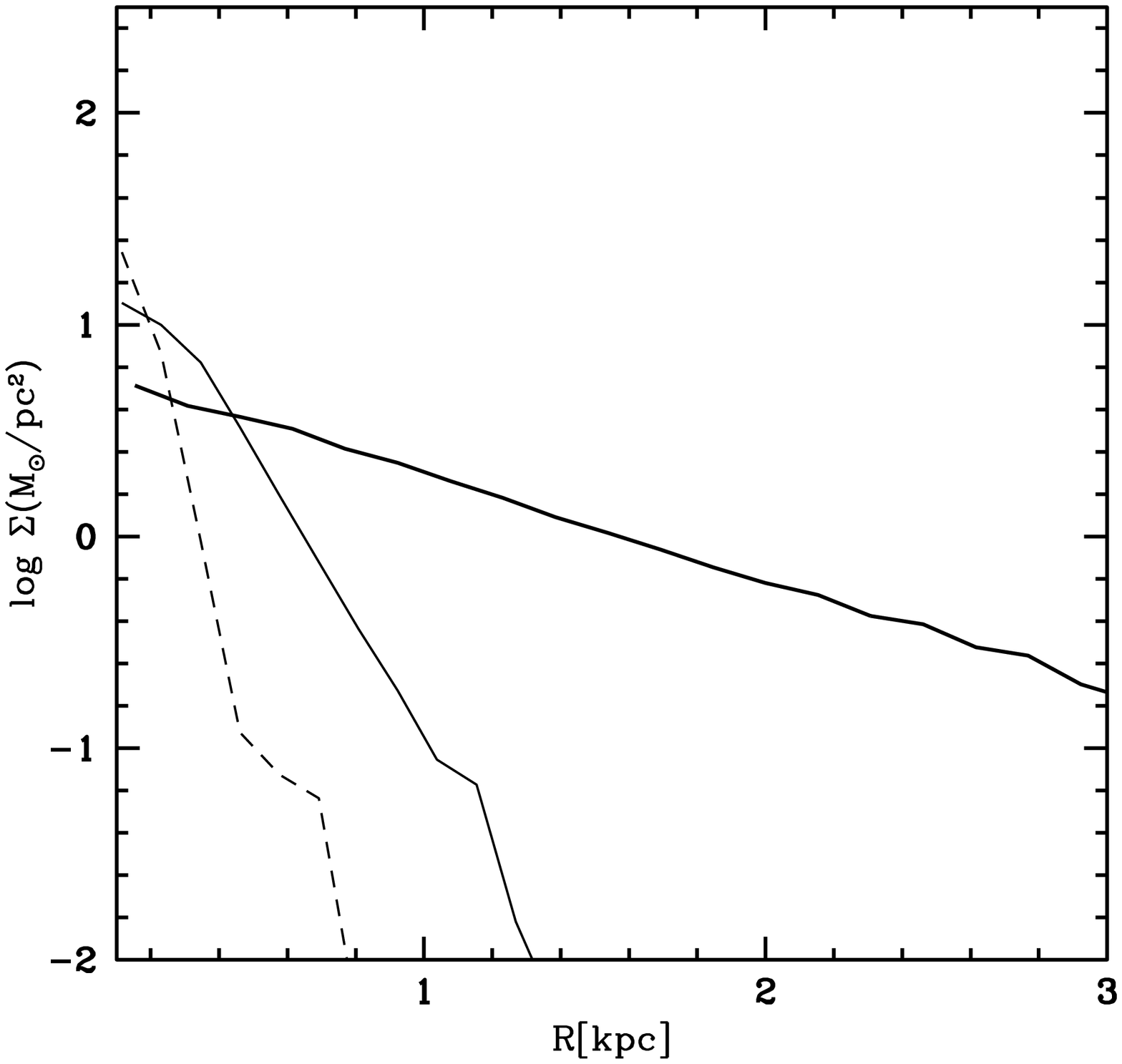}
\epsfxsize=4.5truecm
\hskip 0.4truecm
\epsfbox{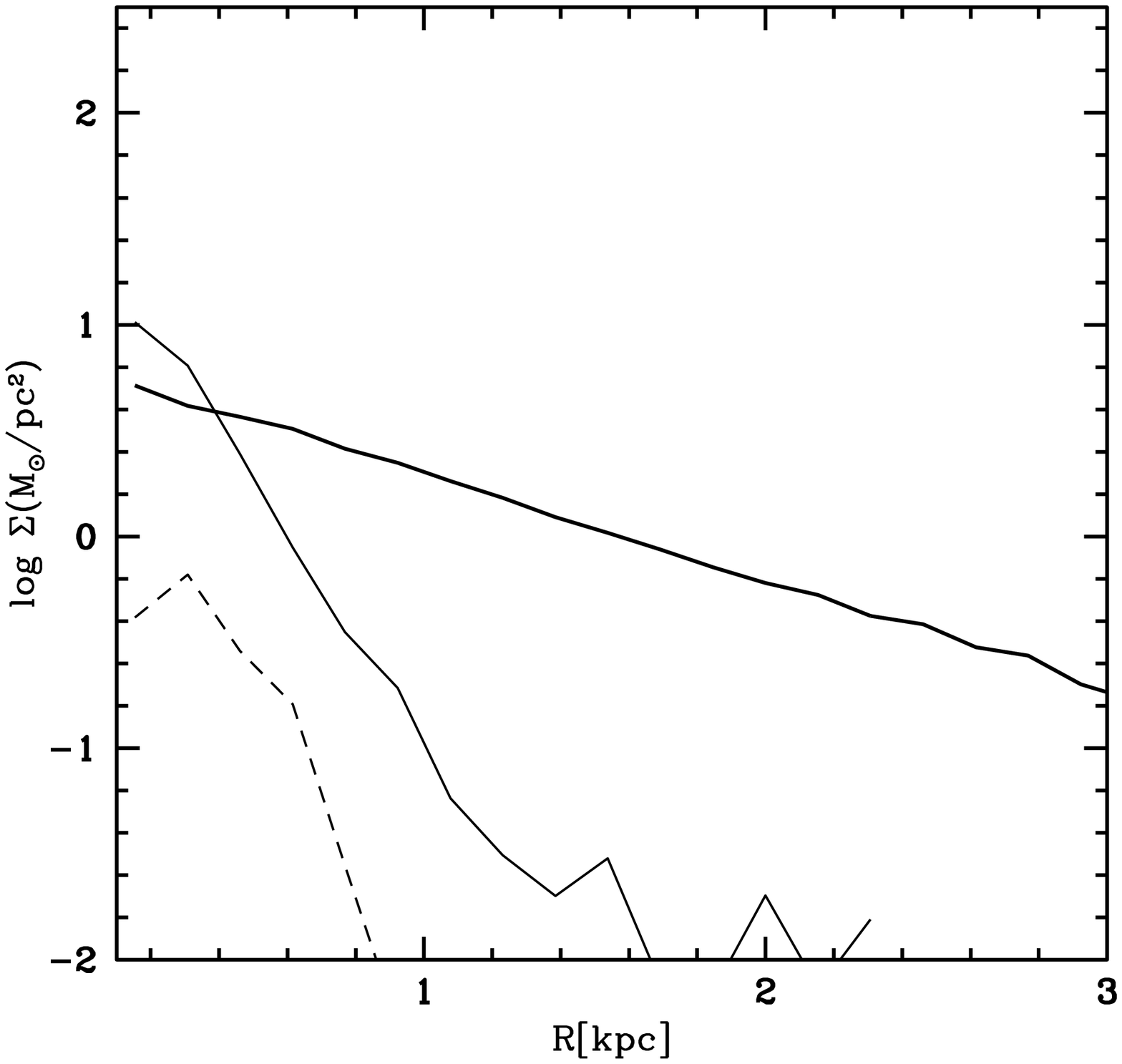}
%\epsfxsize=6truecm
%\epsfbox{v75c4UVsurfgas.ps}
\caption{Evolution of gas surface density profiles in FI runs. The thick solid 
line is used for the initial profile, the thin solid
line for the profile after first pericenter crossing (T=1 Gyr) 
and the dashed line for the profile after second pericenter passage (T=3
Gyr). From left to right, run Rc20TD, Rc20RC, and RC20RCUV are shown.}
\end{figure*}

Individual clouds survive for
a timescale of $\sim 10^8$ years.
The dissolution of the clouds is due to a combination of ram pressure
and tidal disruption, but spurious evaporation due to artificial viscosity 
and discreteness effects 
(particle-particle collisions) may be important,
especially for the smallest clouds that are close to the resolution limit
(we consider a cloud to be unresolved if it contains less than 64 particles,
i.e. twice the number of neighboring particles used in the SPH calculation). 
We calculate that ablation of the clouds due to Kelvin-Helmoltz
stripping would occur on timescales 
comparable to the dissolution timescales observed in the simulations.
%Overall numerous blobs are present
%for the duration of the first orbit, from 1 to 2 Gyr after ram pressure 
%stripping begins. 
%If several interacting dwarfs were simultaneously present
%one might expect to a large number of such blobs around the main 
%galaxy at any given time since their survival times is comparable to the 
%the orbital time at a few tens of kiloparsecs from the center.

A single stripped satellite produces tens of blobs with masses
in the range $10^4-10^6 M_{\odot}$ and sizes in the range $0.5-5$ kpc.
%We caution that the dissolution timescales,
%as well as the final masses and sizes of the clouds, may vary depending on the
%actual cooling rate in such clouds. In particular, our simulations assume
%zero metallicity while a low but non negligible metal content would be
%present in the gas, especially in the case of the largest dwarfs (Mateo 1998).
%Such blobs eventually would fall towards the disk due to drag
%by ram pressure.
These blobs 
could explain the cold HI fragments identified by Thilker et al. (2004)
around M31. These clouds have column densities ($\sim 10^{20}$ cm$^{-2}$) 
and masses (if placed at roughly the distance of M31) of $10^5-10^6 
M_{\odot}$, similar
to  the most massive among our clouds. They have different 
velocities with respect to the systemic velocities, which lead Thilker
et al. to suggest that they could hardly be tidal debris of disrupted
dwarfs. However, a spread in velocities arises naturally because of the
drag exerted by the diffuse gaseous halo.
We measure velocities differing by up to a factor of 2 among the clouds
in our simulations. 
%clouds at the tip of
%the tail suffer stronger drag and are decelerated more since they are in
%a denser part of the halo compared to filaments upstream along the tail.
The variety of shapes of the clouds detected by Thilker et al. 
,from nearly spherical to elongated filaments, is
also reproduced by our simulations (Figure 17).
The clouds do not appear tidally
disturbed because they are  pressure confined.
However, their origin is partly tidal since they
are the result of the combined removal of gas by tides and ram pressure.
Although our simulations lack low temperature cooling (by molecules and
metals) the heating from the cosmic or local UV radiation should keep
their temperature and densities close to the values observed in the
simulations. However, since many of the clouds are just above
our resolution limit  we feel that it is premature
to draw firm conclusions concerning their structure.
Nonetheless the simulations strongly suggest
a connection with HI fragments and possibly other gaseous structures observed
in the Local Group like some of the high velocity clouds (HVCs) and the
Magellanic Stream (see also Mastropietro et al. 2004).

%Part of the HVCs population might be explained by them as well. Metal
%cooling would render the blobs even more resilient to disruption and probably
%more compact. While the masses that we estimate are reasonable, the sizes
%might thus be underestimated since we lack metal cooling.

\begin{figure*}
\hskip 1.1truecm
\epsfxsize=9truecm \epsfbox{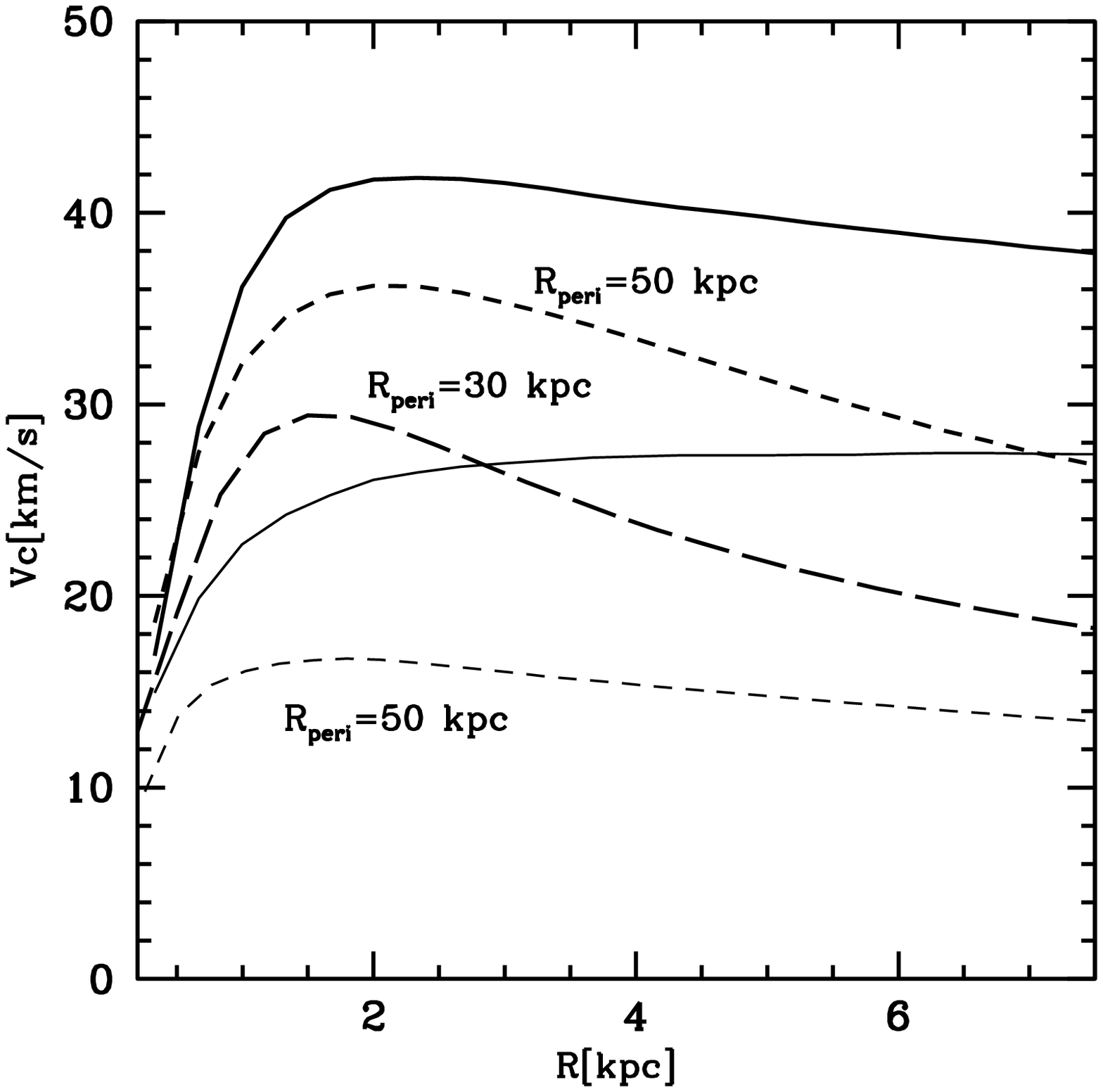}
\caption{Evolution of the rotation curves of the dwarfs 
(including the contributions of both dark matter and baryons)  
in run Rc20TD  (thick lines)and run Rc4aTD (thin lines). The
solid lines represent the initial conditions while the short-dashed show
the curves after 3 orbits, corresponding to 
$t=4$ Gyr and $t=10.2$ Gyr, respectively. The thick long-dashed line shows the
curve for run Rc20bAD after $t=10.2$ Gyr. Pericenter distances for the
various orbits are indicated in the plot. Note that runs Rc20TD and runs
Rc20bAD employ the same initial dwarf model but different orbits.}
\end{figure*}

\section{Summary and Conclusions}

The combination of ram pressure and tidal forces can produce systems
resembling gas poor dwarf spheroidals from  disk-like progenitors.
Here we summarize our findings:

\begin{itemize}

\item
Ram pressure alone can remove completely the gas only in dwarfs having halos
with $V_{peak} \simlt 30$ km/s. This is consistent with the results found
by Marcolini et al. (2003) using different initial conditions and a different
numerical technique.

\item
Compressional heating arising as the dwarf travels through the outer medium
can significantly enhance ram pressure stripping by producing a more 
extended gas distribution. As a result, stripping is more effective in
adiabatic runs than in cooling runs since in the latter such heating is
dissipated.

\item
The combination of tides and ram pressure is in general more effective
at removing gas from the dwarfs than ram pressure alone and always 
significantly more effective than tides alone. Tidal stripping lowers
 $V_{peak}$ and thus the depth of the potential well of the dwarf.

\item 
Tidally induced bar formation opposes ram pressure stripping by driving a 
large fraction of gas towards the center where it is harder to strip at 
subsequent pericenter passages. This counteracting effect is stronger 
when radiative cooling is included and explains why in a few cases the 
addition of tides does not increase the amount of stripped mass.

\item
Low baryonic contents and high mass-to-light ratios comparable to those
of many dSphs can be obtained as a result of ram pressure and tidal stripping.
However heating sources,
such as the cosmic UV background arising during reionization, are required
in order to obtain a very low gas fraction in a halo as massive as that in
which Draco is probably embedded.

%\item
%The decreasing gas content, earlier truncation of star formation history
%and higher mass-to-light ratios of dSphs and dIrrs/dSphs with decreasing 
%galactocentric distances can be explained as a result of infall of similar
%progenitors at different cosmic epochs. In fact the characterictic orbital
%radius is expected to scale with the virial radius of the primary system,
%which scales close to linearly with redshift, and strength of both ram 
%pressure and tidal shocks will scale with the mean orbital radius. 

\item
The gaseous stream produced by the stripped dwarfs fragments into several
cold clouds pressure confined by the outer medium. These clouds have
properties reminiscent of gaseous structures possibly
associated with the Local Group, including some of the HVCs and the
HI fragments recently detected around M31.

\end{itemize}

We have considered only two types of orbits in this paper. Since proper
motions of dSphs are poorly constrained (recent HST-based measurement still
have a quite large error, e.g. see Piatek et al. 2003 for Carina)
there is still quite some freedom
in choosing the initial orbits. For example evolving model V40c20 on an
orbit with smaller pericenter, say between 10 and 20 kpc, would 
enhance both tidal stripping and 
ram pressure.  More eccentric orbits would lead to higher
pericentric velocities which greatly increases the efficiency of 
ram pressure stripping.  Furthermore, the progenitors
of dSphs might have suffered
both tidal and ram pressure stripping already before entering the Milky Way 
halo (Kravtsov et al. 2004).
Hence our results should be regarded as quite conservative.

As we noted above, the initial orbits and initial masses of the dwarfs are 
both factors that the determine the final 
dark-to-baryonic mass ratio and other properties of the remnants. Therefore,
the fact that Fornax and Draco have similar masses, as
inferred from their velocity dispersions (Kazantzidis et al. 2004), but
differ by more than a factor of $10$ in their luminosity, can be explained 
in two ways. Either the progenitors of these two dwarfs started out with
very different relative amounts of dark matter and baryons, for reasons
related to their formation history and not to the environment, or they
descend from very similar progenitors having comparable dark and baryonic
masses, but
evolved differently because they entered the primary halo at a different
epoch and on different orbits. 
%In the second case it is the different 
%environment they experimented, namely different pericenter distances 
%and orbital times, that sets the different evolutionary 
%history.  
The ionizing radiation at high redshift also plays a role in 
determining the efficiency of ram pressure stripping, and its effect
again will be stronger for dwarfs infalling at higher redshift.
It is a natural consequence of structure formation in hierarchical
models that subhaloes that are found closer to the center of the primary
system at $z=0$ are typically those that fell earlier into the potential well
of the primary system. 
Based on their current distances 
one would conclude, for example, that
Fornax is likely a more recent addition to the
Milky Way system, and would have suffered fewer and weaker pericenter
passages, relative to Draco. It retained some gas after the first one
or two orbits, a condition usually satisfied in those of our simulations 
that have the largest pericenter, and underwent a few bursts over the past few 
Gyr (Mayer et al. 2001b). Instead, Draco entered earlier on a tighter
orbit and lost almost all its gas at the first pericenter passage, 
thus ending star
formation early. We note that for the orbits with a pericenter of $30$ kpc
pericenter crossing occurs $0.7$ Gyr after the dwarf enters the Milky Way
halo. If the dwarf begins its infall at $z=6-7$, i.e 
$\sim 1$ Gyr after the Big Bang, this leaves about $\sim 2$ 
Gyr to form stars before gas removal occurs, consistent with the estimated
time span for the star formation history of Draco (Grebel \& Gallagher 2004).
While the infall time of the progenitor of Draco is clearly a free parameter,
our results show that these simple assumptions produce a scenario that fits a 
number of observational constraints.
%In addition, we
%saw that the ionizing radiation at high redshift also plays a role in 
%determining the efficiency of ram pressure stripping, and its effect
%again will be stronger for dwarfs infalling at higher redshift.

\begin{figure*}
\hskip 1.1truecm
\epsfxsize=10truecm 
\epsfbox{clouds1.ps4}
\end{figure*}
\begin{figure*}
\hskip 1.1truecm
\epsfxsize=10truecm 
\epsfbox{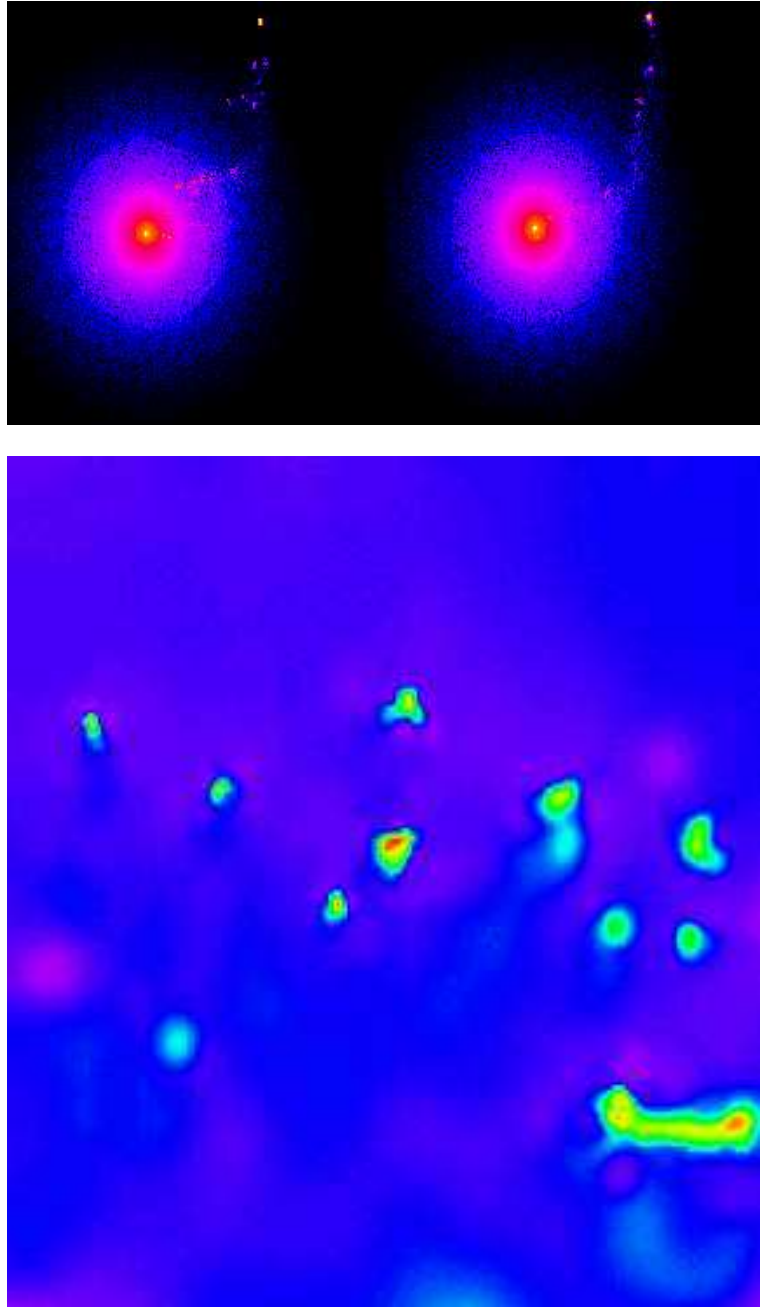}
\caption{Top: color coded density map (brighter colors for higher densities) 
showing a box of 200 kpc on a side, slightly offset from the center of mass of 
the Milky Way+satellite system, for run Rc20RC (left) and run  Rc20RCUV
(right) at $t=1.5$ Gyr. A gaseous trail fragmenting into clouds is visible.
Bottom: color coded temperature map of one small region of the gaseous stream
in run Rc20RC (the box has a size of about 40 kpc on a side) showing the
detailed structure of the clouds. }
\end{figure*}

It is important to keep in mind that the Milky Way halo model
used here is based on  data at $z=0$ and the
components that we assumed as fixed do indeed evolve through cosmic
time. At high $z$,
as the dwarfs began their route towards the Milky Way, the
dark halo and gaseous halo masses may have been quite
different. However, $\Lambda$CDM galaxy
formation simulations suggest that large disk galaxies had the last major
merger quite early on in order for a large enough disk to assemble
by $z=0$, which means most of the dark halo mass was probably already in
place at $z=2-3$. The evolution of the gaseous halo in $\Lambda$CDM simulations
has not been investigated yet but work in progress
suggests its density at several tens of kiloparsecs could 
have been ten times higher than the value adopted in this paper. 
%(Governato, Mayer et al., in preparation). 
%Therefore our
%results on the effectiveness of combined ram pressure and tidal stripping
%Some of the progenitors
%of dSphs might have suffered
%both tidal and ram pressure stripping already before entering the Milky Way 
%halo (Kravtsov et al. 2004). Therefore, our estimates
%on the total mass losses from the combined effect of ram pressure and
%tides have probably to be regarded as lower limits.

In our simulations ram pressure stripping is most effective
close to pericentric passages. This reflects the fact that we have
assumed a smooth density distribution for the hot halo. If the halo
has a complex multiphase structure as predicted by some models
(Maller \& Bullock 2004) the dwarfs would move through a clumpy
medium and could undergo intense ram pressure stripping even far
from pericenter if they encounter a dense and massive  
cloud. Galaxy formation simulations which follow the build-up
and evolution of the gaseous halo at very high resolution with 
a proper treatment of the multi-phase ISM will eventually
be able to address this aspect. However, current 
cosmological simulations can hardly capture the effect of ram pressure 
on satellite galaxies even in the case of a smooth halo
because of the limited resolution and
owing to uncertainties in the star formation and feedback processes
which concur to determine the gas fraction in satellite galaxies.
%This is not only because of the limited 
%resolution of the SPH component of such simulations, but also because all 
%processes to which the gas component participates, from the conversion into 
%stars to heating by various forms of feedback, are highly  uncertain both
%from a physical and from a numerical point of view. The numerical limitations
%are especially severe at the scale of dSphs, below 30 km/s, that are resolved 
%by only a few hundred particles in even the highest resolution simulations
%published so far (Governato, Mayer et al. 2004). One problem, in particular,
%is that, even with the inclusion of supernovae feedback, the conversion of 
%gas into stars tends to be too efficient at high $z$ , with the result that 
%satellites enter the primary halo already with very low gas fractions,
%so that ram pressure cannot affect them significantly and they end up being
%too bright at $z=0$ (Governato, Mayer et al., in preparation). 
%Numerical loss of angular momentum of SPH particles in the poorly resolved 
%satellites which artificially
%enhances star formation since the gas becomes very centrally concentrated 
%(Kauffmann, Mayer et al. 2004) and the lack of additional heating sources
%at high $z$, for example Population III stars, are two possible reasons
%behind this discreapancy. 
Therefore nowadays simulations such as those discussed in this paper
represent a necessary counterpart to cosmological runs and provide
the only means to test the physical effect of environmental mechanisms in a
robust way.

%{\bf ACKNOWLEDGMENTS}

\vskip 1truecm

The numerical simulations were carried out on the zBox supercomputer at the
University of Zurich and on LeMieux at the Pittsburgh Supercomputing Center.

{}

\end{document}